\documentclass[11pt]{article}

\usepackage[final]{acl}

\usepackage{times}
\usepackage{latexsym}
\usepackage[T1]{fontenc}

\usepackage[utf8]{inputenc}

\usepackage{microtype}

\usepackage{inconsolata}

\usepackage{graphicx}

\usepackage{pgfplots}
\usepackage{pgfplotstable}
\usepackage{sansmath}
\usepackage{helvet}
\usepackage{graphicx}
\usepackage{amsmath}
\usepackage{amssymb}
\usepackage{float}
\usepackage{tcolorbox}
\usepackage{etoc}
\usepackage{listings}
\usepackage{scrextend}
\usepackage{subcaption} 
\usepackage{xspace}
\usepackage{multirow}
\usepackage{wrapfig}
\usepackage{titletoc}
\usepackage{booktabs}
\usepackage{colortbl}
\usepackage{xcolor}
\usepackage[table]{xcolor} 
\newcommand{\huggingface}{\raisebox{-2pt}{\includegraphics[height=1.3em]{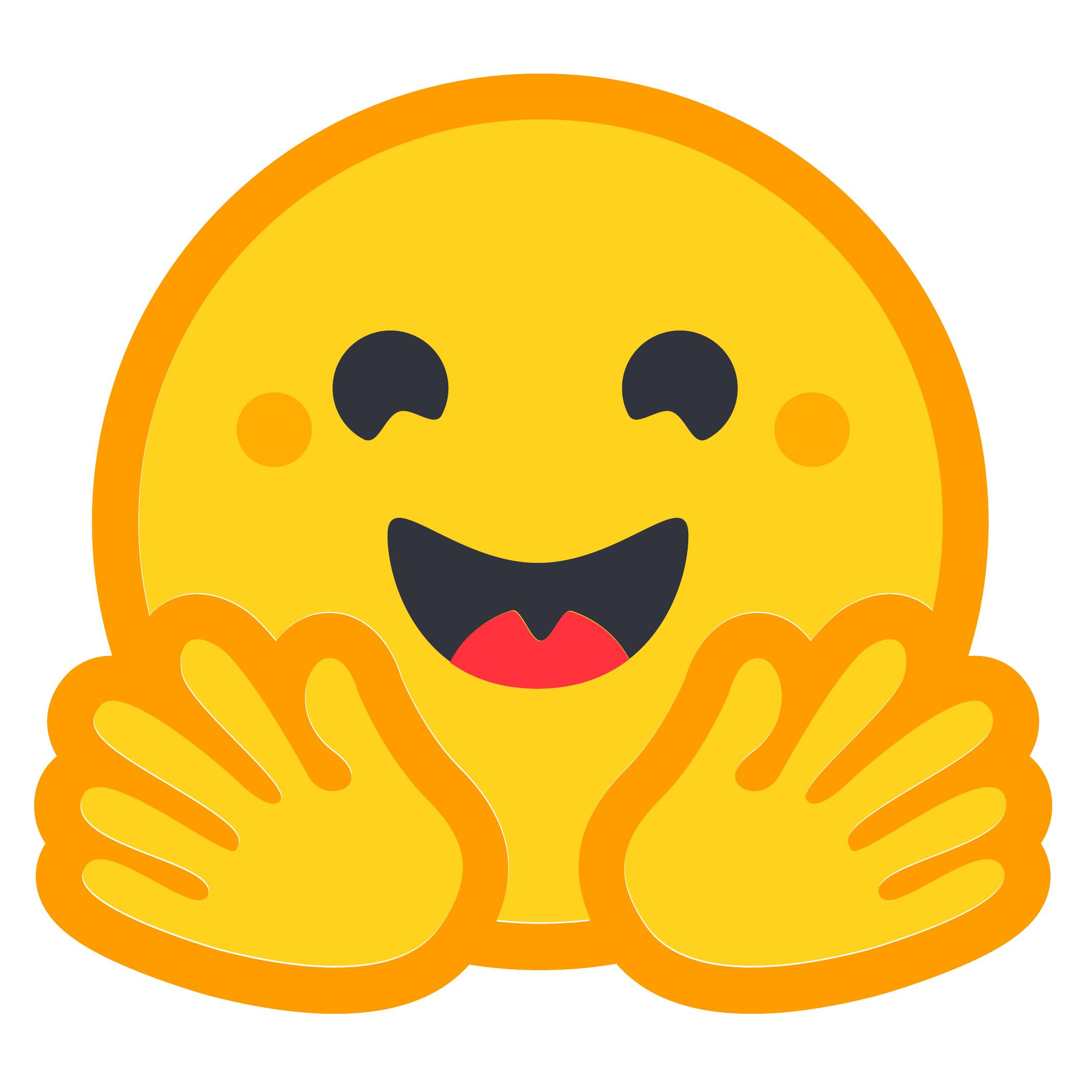}}\xspace}
\newcommand{\github}{\raisebox{-2pt}{\includegraphics[height=1.3em]{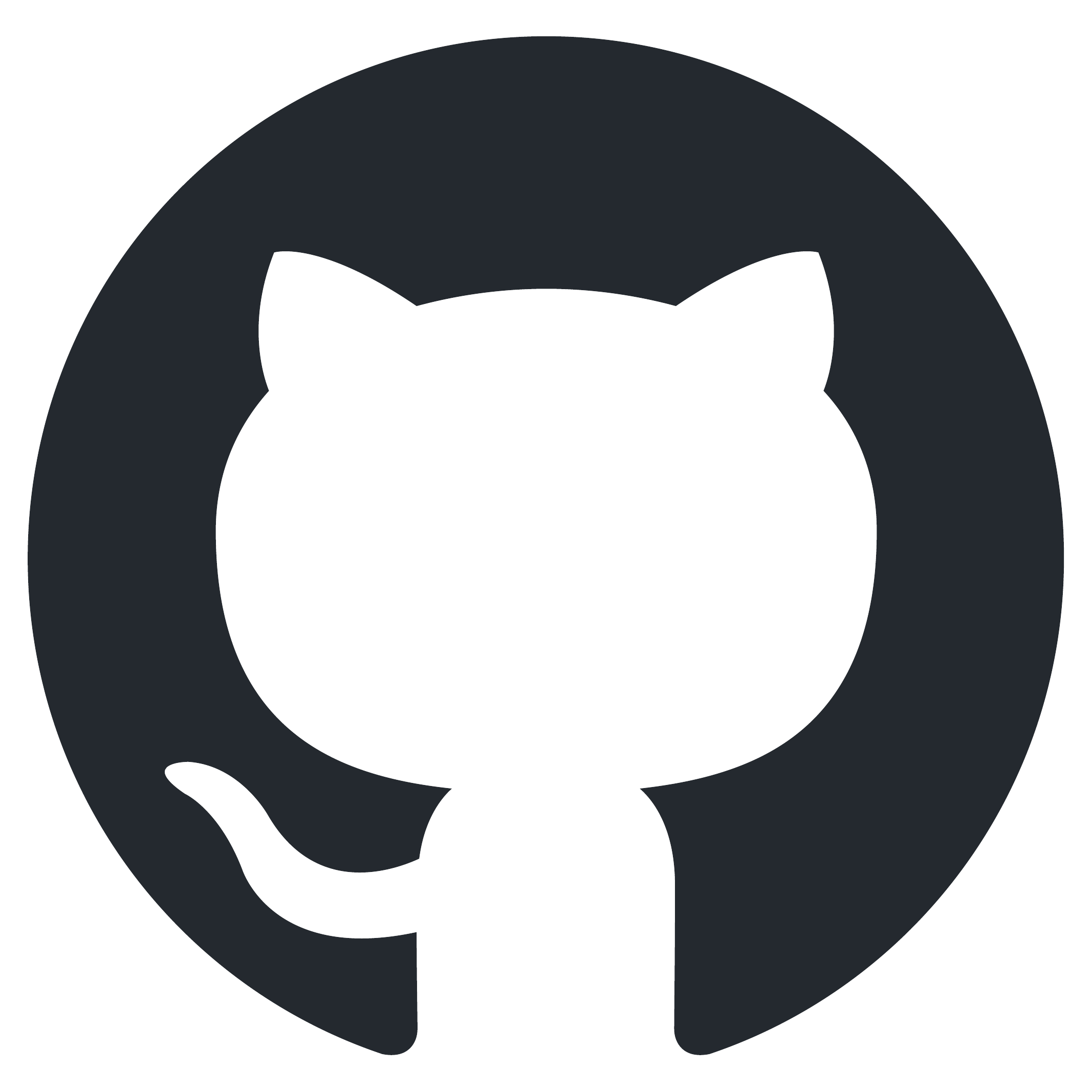}}\xspace}

\newlength\mystoreparindent

\definecolor{sectiongray}{gray}{0.9}

\newcommand{\sizhe}[1]{\textcolor{black}{#1}}

\title{CodeFlowBench: A Multi-turn, Iterative Benchmark \\for Complex Code Generation}

\author{
    \textbf{Sizhe Wang}$^\dagger$\textsuperscript{1,3},
    \textbf{Zhengren Wang}$^\dagger$\textsuperscript{1,2},
    \textbf{Dongsheng Ma}\textsuperscript{1}\\
    \textbf{Yongan Yu}\textsuperscript{\textbf{1,4}}, 
    \textbf{Rui Ling}\textsuperscript{\textbf{1}},
    \textbf{Zhiyu Li}\textsuperscript{*\textbf{2}},
    \textbf{Feiyu Xiong}\textsuperscript{\textbf{2}},
    \textbf{Wentao Zhang}\textsuperscript{*\textbf{1,5,6}}\\
    \textsuperscript{1}Peking University
    \textsuperscript{2}Institute for Advanced Algorithms Research, Shanghai \\
    \textsuperscript{3}Shanghai University of Finance and Economics
    \textsuperscript{4}McGill University
    \textsuperscript{5}Zhongguancun Academy\\
    \textsuperscript{6}Beijing Key Laboratory of Data Intelligence and Security (Peking University)\\
    \texttt{\{lizy,xiongfy\}@iaar.ac.cn}, 
    \texttt{wentao.zhang@pku.edu.cn}\\
\vspace{.5em}\huggingface \href{https://huggingface.co/datasets/WaterWang-001/CodeFlowBench-2505}{Dataset}
\hspace{0.6em}
\vspace{.5em}\github \href{https://github.com/Rise-1210/codeflow}{Code}
}

\begin{document}

\maketitle
\deffootnote[1.5em]{1.5em}{1em}{}
\renewcommand{\thefootnote}{\fnsymbol{footnote}}
\footnotetext{$\dagger$ Equal contribution;  * Corresponding author. \\ The first author completed this work during an internship at Peking University.
}
\renewcommand{\thefootnote}{\arabic{footnote}}
\begin{abstract}
Modern software development demands code that is maintainable, testable, and scalable by organizing the implementation into modular components with iterative reuse of existing codes. We formalize this iterative, multi-turn paradigm as \textit{codeflow} and introduce \textbf{CodeFlowBench}, the first benchmark designed to comprehensively evaluate LLMs’ ability to perform codeflow - implementing new functionality by reusing existing functions over multiple turns. CodeFlowBench comprises two complementary components: CodeFlowBench-Comp, a core collection of 5,000+ competitive programming problems from Codeforces updated via an automated pipeline and CodeFlowBench-Repo, which is sourced from GitHub repositories to better reflect real-world scenarios. Furthermore, a novel evaluation framework featured dual assessment protocol and structural metrics derived from dependency trees is introduced. Extensive experiments reveal significant performance degradation in multi-turn codeflow scenarios. Furthermore, our in-depth  analysis illustrates that model performance inversely correlates with dependency complexity. These findings not only highlight the critical challenges for supporting real-world workflows, but also establish CodeFlowBench as an essential tool for advancing code generation research.

\end{abstract}

\section{Introduction}
Large Language Models (LLMs) have revolutionized code generation, with benchmarks like HumanEval \citep{chen2021evaluating} and MBPP \citep{austin2021program} establishing foundational standards. As LLM capabilities advance, their role in real-world software development has expanded beyond solving toy problems to supporting complex workflows \citep{jiang2024survey,jin2024llms,liu2024large}. Modern benchmarks such as SWE-Bench \citep{jimenez2023swe,li2024devbench} now emphasize practical scenarios like bug fixing. However, current benchmarks \citep{gu2024cruxeval,huang2024effibench,wang2025maintaincoder} still overlook the critical aspect of real-world development: the multi-turn and iterative \textit{codeflow} scenario.

\begin{figure}[tb]
\centering
\includegraphics[width=0.9\linewidth]{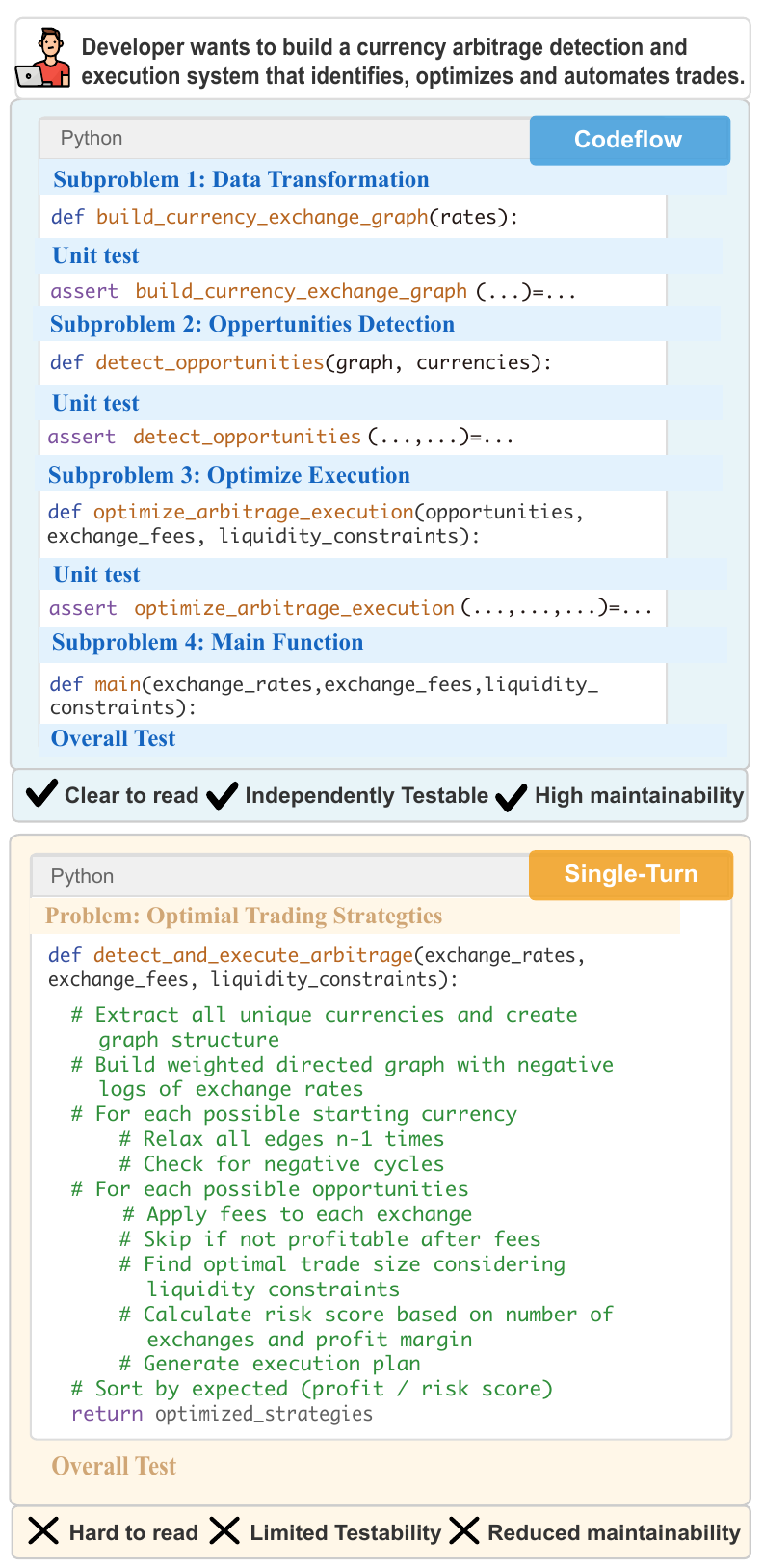}
\caption{A currency arbitrage example contrasting the
modular \textit{codeflow} workflow (top) with a monolithic
single-turn implementation (bottom).}
\label{codeflow}
\end{figure}

\paragraph{The CodeFlow Task} In modern software engineering, \sizhe{workflows based on \textit{iterative implementation}} are becoming increasingly prevalent, as the cornerstones of best practices like agile development \citep{larman2004agile,abrahamsson2017agile,cram2019agile}. \sizhe{Unlike \textit{iterative refinement}, this paradigm requires} breaking down complex tasks into manageable subproblems, progressively \sizhe{building} solutions, and \sizhe{reusing} modular functions. \sizhe{Through this sequential approach,} developers can achieve faster delivery, reduced redundancy, and enhanced maintainability in teamwork \citep{haefliger2008code,feitosa2020code}. 
For example, React's core package alone sees over 37 million weekly downloads across 2,300+ dependent modules, illustrating the productivity gains of modular reuse \cite{boduch2019react}. As shown in Fig. \ref{codeflow} and \ref{codeflowbench}, by building solutions via traversing the function dependency tree from the bottom up, the structured \textit{codeflow} can both enables parallel development and improves readability, testability and maintainability through modular and responsibility boundaries. To integrate effectively into codeflows, LLMs must learn to ``look before and after'', namely accurately leveraging pre-implemented functions while ensuring the modularity of codes to generate.

\paragraph{Motivation} Despite the growing demands, current benchmarks have not fully captured the multi-turn and iterative aspects of codeflow. Firstly, most benchmarks, such as HumanEval and MBPP, focus only on single-turn code generation. While recent benchmarks like BigCodeBench \citep{zhuo2024bigcodebench} and SWE-Bench \citep{jimenez2023swe} have begun to incorporate practical development scenarios, they still remain at the level of single-turn code modification (or two-turn code generation), leaving the multi-turn generation capabilities unknown. Secondly, the few existing multi-turn benchmarks such as MTPB \citep{nijkamp2022codegen}, only focus on single-function programming, lacking both unit tests and sufficient complexity to reflect real-world dependencies. Finally, due to the absence of an update mechanism, previous static datasets risk contamination and unreliable assessment. 
Therefore, there is a pressing need for a more challenging, well-annotated and frequently updated benchmark specially designed for codeflow.

To bridge this gap, we introduce CodeFlowBench, a benchmark suite that integrates two complementary datasets: \textit{CodeFlowBench-Comp}, which is sourced from competitive programming platforms like Codeforces\footnote{\url{https://codeforces.com/}} and \textit{CodeFlowBench-Repo}, which is derived from diverse real-world GitHub repositories via DomainEval \citep{zhu2025domaineval}. Utilizing a unified automated pipeline, CodeFlowBench transforms raw data into sequences of multi-turn, iterative subproblems, accompanied by verified solutions and test cases. This dual-component architecture ensures a broad evaluative scope, spanning reasoning-intensive algorithmic challenge and labor-intensive software engineering. Specifically, \textit{CodeFlowBench-Comp} provides high-difficulty, high-quality problems with periodic updates, while \textit{CodeFlowBench-Repo} injects real-world complexity into the assessment. To further reveal model deficiencies, CodeFlowBench introduces a specialized evaluation framework and incorporates structural labels and metrics derived from dependency trees, enabling a nuanced analysis of multi-turn performance.

\paragraph{Contributions} Our contributions are threefold:
\begin{itemize}
 \item \textbf{Pipeline Innovation}: We developed a universal data curation pipeline grounded in function dependency analysis. This fully-automated and lightweight framework decomposes monolithic solutions into multi-turn, iterative coding problems that necessitate strategic code reuse. Its design is rigorously verifiable and highly extensible, enabling the transformation of diverse source code into structured codeflow-style challenges.

\item \textbf{Benchmark Construction}: We introduce CodeFlowBench, the first benchmark to evaluate the iterative, multi-turn code generation capabilities of LLMs. By integrating two complementary datasets: \textit{CodeFlowBench-Comp} (algorithmic depth) and \textit{CodeFlowBench-Repo} (real-world domain breadth), we provide a comprehensive testbed to examine software engineering proficiency.

\item \textbf{Evaluation Design \& Insights}: 
We propose a novel evaluation framework that directly contrasts multi-turn and single-turn generation patterns. 
By introducing structural metrics, such as Average Pass Depth (APD) and Dependency Structure Complexity (DSC), derived from dependency trees, our framework captures the unique complexities of iterative tasks. Extensive experiments reveal a significant performance degradation in codeflow scenarios, even among state-of-the-art reasoning models, highlighting a critical frontier for future research and the need for more advanced codeflow programming capabilities.

\end{itemize}

\section{Related Work}

\paragraph{Code Generation Benchmarks}
The landscape of code generation benchmarks has evolved from simpler to more complex tasks to keep pace with the rapid development of LLMs \citep{li2022competition, quan2025codeelo, yu2024humaneval}, but still fail to comprehensively capture the multi-turn and iterative features of real-world scenarios. Early works like HumanEval \citep{chen2021evaluating} and MBPP \citep{austin2021program} focus on  standalone functions with low complexity and limited dependency environments. Recent benchmarks have emerged to evaluate more complex and realistic scenarios, yet have obvious limitations: most benchmarks such as APPS \citep{hendrycks2021measuring}, LiveCodeBench \citep{jain2024livecodebench} and SWE-Bench \citep{jimenez2023swe} are limited to single-turn code generation or modification. For the few existing multi‑turn benchmarks, MTPB \citep{nijkamp2022codegen} focus on overly simplistic single-function programming without paired unit tests, while InterCode \citep{yang2023intercode} discusses interactive coding with execution feedback. In stark contrast, \textit{codeflow} structures the development into multi-turn processes, ensuring each component is maintainable, testable and reusable. These limitations highlight the crucial gaps for codeflow benchmarking, and the CodeFlowBench pioneers this research line.

\paragraph{Code Generation LLMs}
Recent years have witnessed unprecedented progress in code generation capabilities of LLMs. Early works such as Codex \citep{chen2021evaluating} and AlphaCode \citep{li2022competition} demonstrated  proficiency in tasks ranging from code completion to competition-level problem solving. With the scaling up of pre‑trained models, exemplified by GPT \citep{gpt5}, Claude \citep{claude4}, Deepseek‑Coder \citep{guo2024deepseek} and Qwen‑Coder \citep{yang2025qwen3}, these advanced models have impressive performance across various programming tasks, languages and domains. Building on these foundations, the code generation capabilities have further advanced through instruction tuning and agent frameworks. Models such as WizardCoder \citep{luo2023wizardcoder} and Magicoder \citep{wei2023magicoder} leverage instruction tuning to improve intent alignment and interactive dialogue capabilities, while agent frameworks like AgentCoder \citep{huang2023agentcoder} and MapCoder \citep{islam2024mapcoder} enable autonomous planning, iterative refinement, and self-evaluation. Despite advancements, the community still remains unknown about \textit{"how well and how deeply LLMs can perform codeflow"}—a critical paradigm for real-world software engineering. Our CodeFlowBench thus provides a principled framework for advancing both model development and evaluation.

\section{CodeFlowBench}
\begin{figure}[t]
    \centering
    \includegraphics[width=0.9\linewidth]{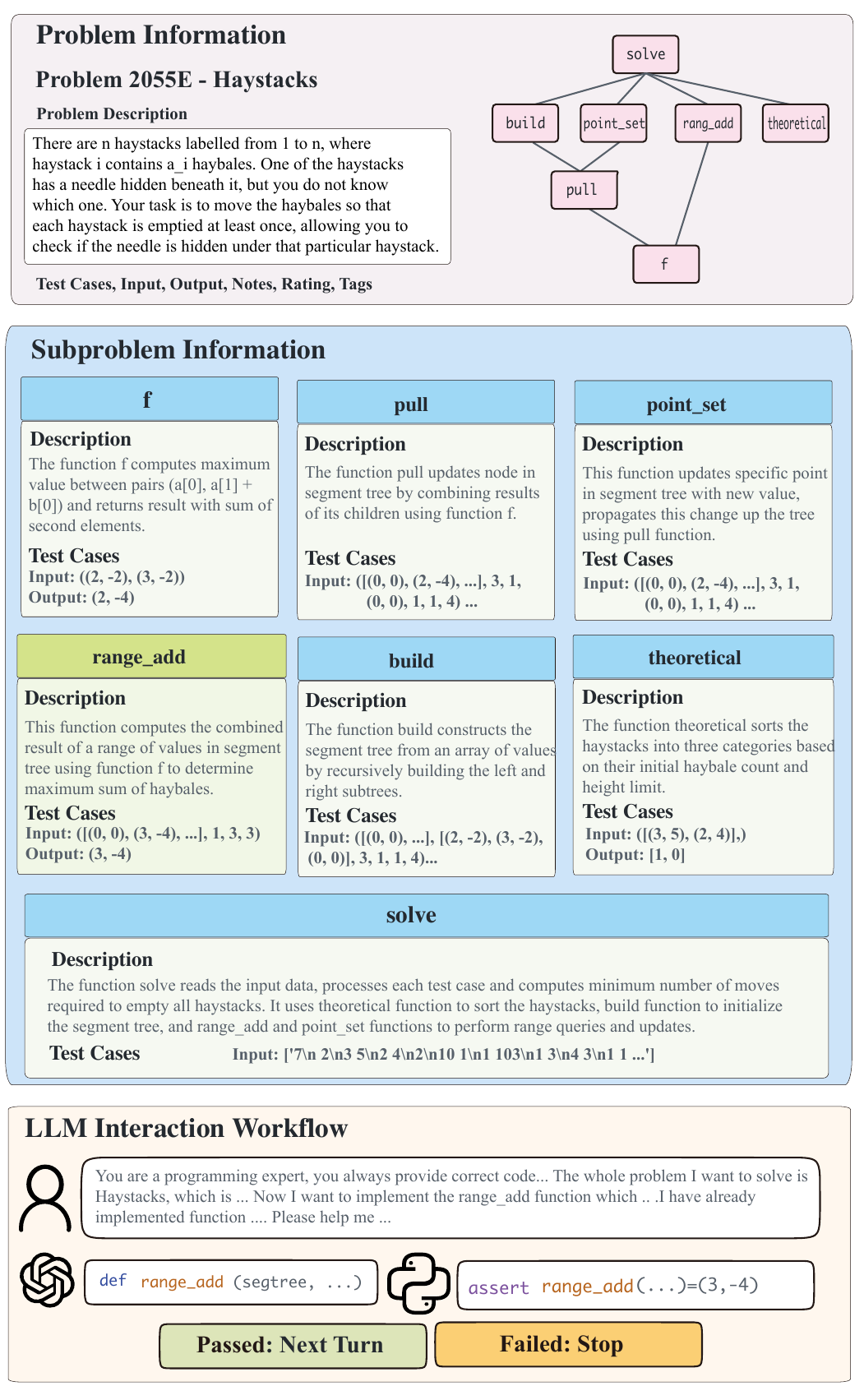}
    \caption{An illustrative example from CodeFlowBench (Source: Codeforces \href{https://codeforces.com/problemset/problem/2055/E}{2055E}). The figure illustrates the
problem definition with its dependency structure (top),
the iterative subproblem decomposition (middle), and
the corresponding LLM interaction workflow (bottom).} 
    \label{codeflowbench}
\end{figure}

In this section, we introduce CodeFlowBench's data curation pipeline and evaluation framework. Fig. \ref{codeflowbench} provides an example of CodeFlowBench. 

\subsection{Data Curation Pipeline}
\label{sec: Data Curation Pipeline}
\begin{figure*}
    \centering
    \includegraphics[width=0.9\linewidth]{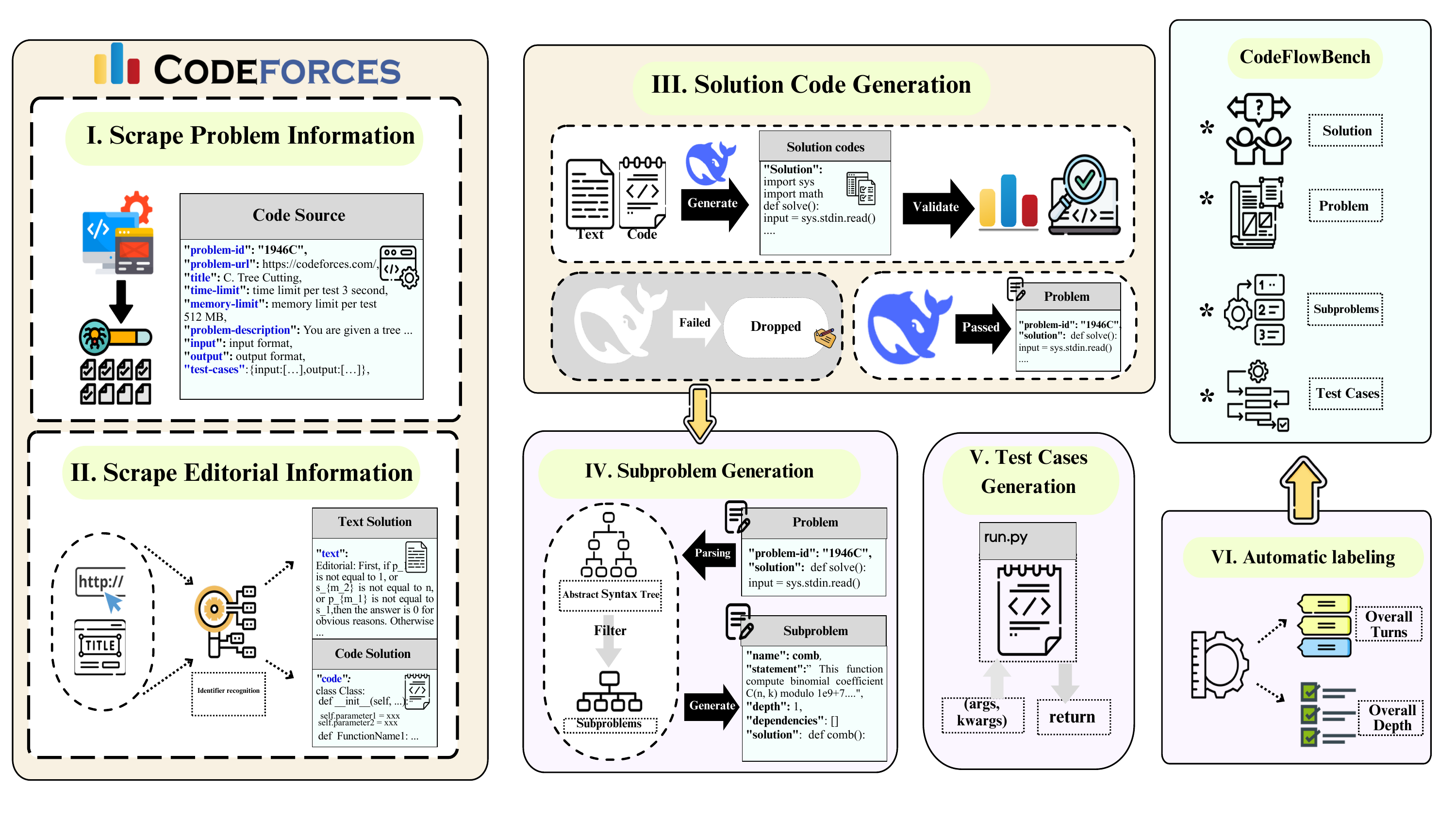}
    \caption{The data curation pipeline of CodeFlowBench. In \textbf{Data Preparation Phase} (Steps I--III), we scrape problem statements and editorial information, followed by generating and validating the canonical solution codes. In \textbf{Subproblem Generation Phase} (Steps IV--VI), we decompose solutions into subproblems via AST parsing, generate corresponding test cases, and perform automatic complexity labeling to achieve the final CodeFlowBench.
    }
    \label{pipeline}
\end{figure*}

We designed an automated, lightweight data-curation pipeline to generate complex multi-turn coding problems. As shown in Fig. \ref{pipeline}, the pipeline mainly consists a data preparation phase and a subproblem generation phase.

\paragraph{Data Preparation Phase}
To address the lack of reasoning-intensive programming datasets tailored for codeflow scenarios, this phase collects programming challenges from Codeforces to construct \textit{CodeFlowBench-Comp}. We first scrape problem metadata and official editorial content from the platform. To ensure data consistency, raw editorials are processed using DeepSeek-R1 \citep{guo2025deepseek} to generate modular, parsable solution code. These generated solutions undergo rigorous verification via the Codeforces submission system; only implementations that pass all official test cases are retained. By anchoring our dataset in official editorials and validating them against a live judge, we guarantee high-quality, correct codeflow implementations for every problem. Further specifics are provided in Appendix \ref{sec: Detail of Problem Scraping}--\ref{sec: Detail of Solution Code Generation}.

\paragraph{Subproblem Generation Phase}
This phase functions as a \textit{general framework} for transforming monolithic solutions into the codeflow format. First, we parse the AST of the verified solution to extract and topologically sort function dependencies. We treat each function as a distinct subproblem, employing Deepseek-V3 \citep{liu2024deepseek} to back-translate its logic into a natural language statement. Next, we execute the full solution against public test cases, instrumenting function calls to capture input-output pairs as ground-truth unit tests for each subproblem. Finally, we annotate problem complexity such as dependency structure complexity, metrics that correlate strongly with official difficulty ratings. Implementation details are provided in Appendix \ref{sec: Detail of Subproblems Generation}--\ref{sec: Detail of Test Cases Generation}.

\textbf{Remark.} Our AST-based decomposition pipeline is fundamentally grounded in established software engineering principles. \sizhe{By decomposing monolithic solutions at the function level via AST parsing, our methodology inherently enforces the Single Responsibility Principle (SRP) and Separation of Concerns (SoC). Each extracted subproblem represents a cohesive logical unit with well-defined interfaces, ensuring the benchmark evaluates realistic modularity.} Moreover, our strict source curation prevents the inheritance of underlying structural flaws. To further ensure this pipeline is not biased by the LLMs used for subproblem description and code generation, we conducted several ablation studies. We assessed description consistency across models, validated quality via Human-LLM cross-validation, and verified the structural stability of code decomposition across different LLMs. Ablation details are found in Appendix \ref{sec: Ablation Study of LLM usage}, and cost estimates for pipeline running are in Appendix \ref{sec: cost estimation}.


\subsection{Benchmark Details}

\begin{figure}[htbp]
    \centering
    \begin{subfigure}[t]{\linewidth}
        \centering
        \includegraphics[width=\linewidth]{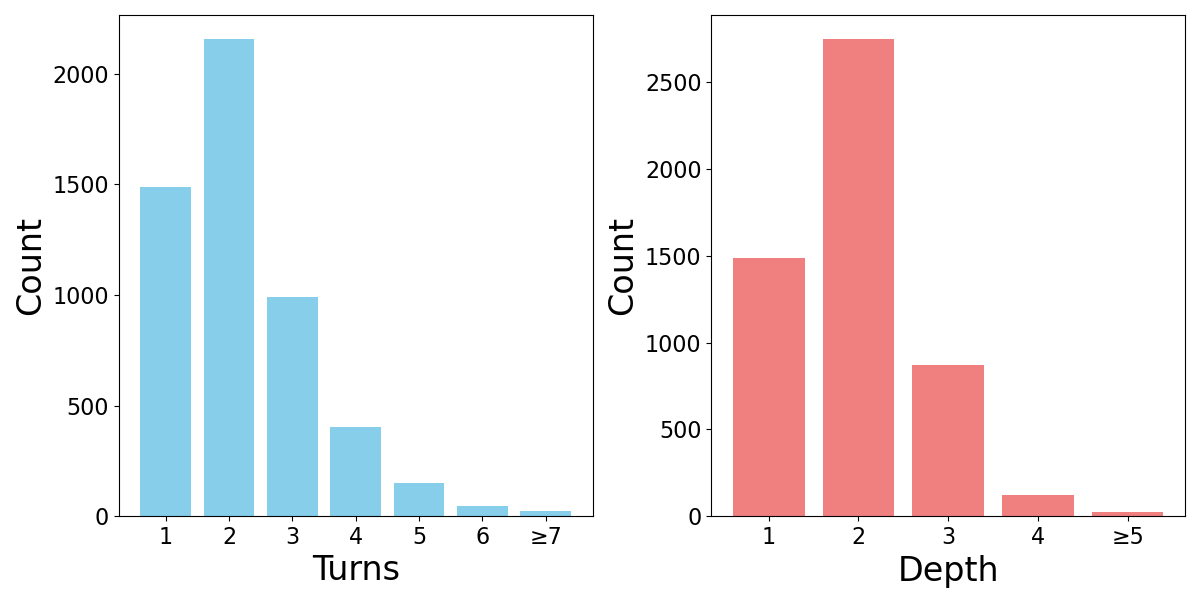}
        \caption{Distributions of overall-turns and overall-depth.}
        \label{fig:distribution}
    \end{subfigure}
    
    \vspace{1em} 
    
    \begin{subfigure}[t]{\linewidth}
        \centering
        \includegraphics[width=\linewidth]{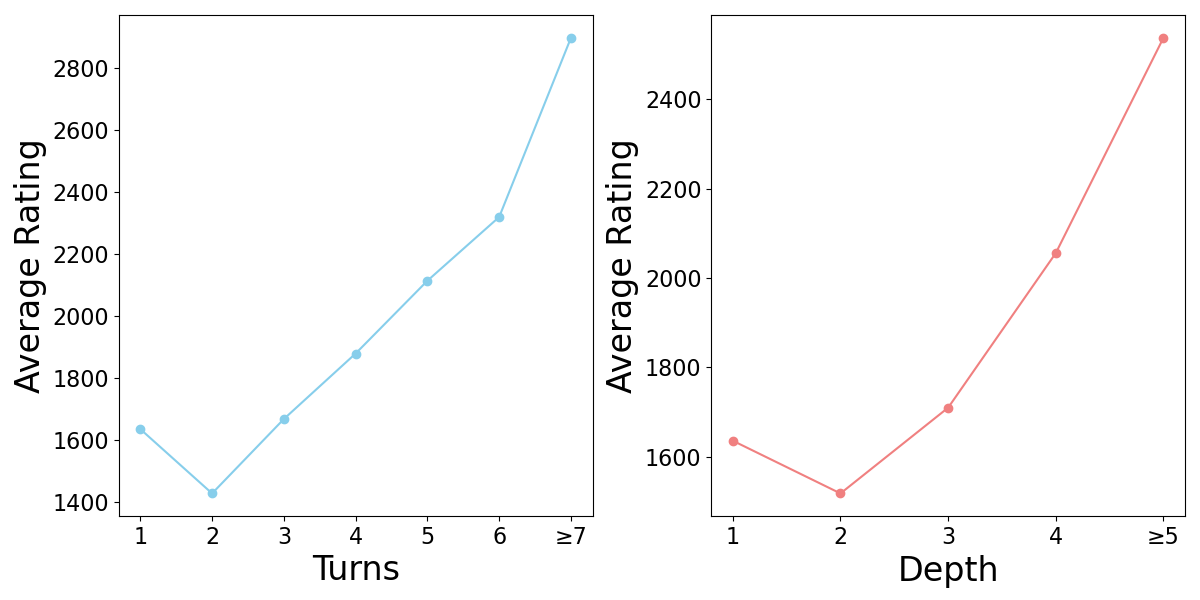}
        \caption{The Correlations with Rating Levels.}
        \label{fig:rating}
    \end{subfigure}
    
    \caption{Statistics of the overall-turns and overall-depth metrics in CodeFlowBench-Comp. Subfigure (b) shows inflection points at turns = 1 and depth = 1. This is attributed to the fact that competition-level problems are not restricted to multi-turn or deeply nested structures. Partial difficult problems are designed to be solvable by single function such as number theory related problems.}
    \label{fig:distribution-rating}
\end{figure}

\paragraph{CodeFlowBench-Comp}
This component comprises 5,258 competitive programming problems designed to test algorithmic reasoning depth. It emphasizes multi-step logic with up to $\ge$7 interaction turns. Similarly, dependency structures are non-trivial, concentrated at depth 2, indicating that nested function calls are the norm rather than the exception. To validate the difficulty of these structures, we correlated our metrics with official Codeforces ratings.
Fig. \ref{fig:distribution-rating} demonstrates the statistical information of CodeFlowBench-Comp. Regarding the distribution, the data shows a clear preference for multi-step logic. Furthermore, correlating these metrics with Codeforces difficulty ratings confirms their validity. As illustrated in the rating curves, we observe a strong positive correlation beyond the structural baseline: as the interaction turns and dependency depth increase, the average problem rating rises sharply—approaching 2,900 for tasks with $\ge$7 turns. This demonstrates that our structural metrics effectively capture the intrinsic complexity of logical problems.

\paragraph{CodeFlowBench-Repo} 
Complementing the algorithmic focus, CodeFlowBench-Repo evaluates software engineering proficiency in real-world contexts. It is constructed by applying our Subproblem Generation pipeline to DomainEval \citep{zhu2025domaineval}, utilizing high-quality GitHub repositories ($>$100 stars) as the source. Our curated dataset spans five specialized domains: \textit{Basic, Cryptography, Network, System, and Visualization}. Through our pipeline, we extended these original contexts into complex codeflow scenarios that reach up to 9 interaction turns and a dependency depth of 4. The final distribution highlights significant structural complexity: 56.6\% of problems require multi-turn iterative generation, forcing models to maintain reasoning continuity, while 54.3\% involve deep dependency chains, requiring the model to correctly orchestrate hierarchical function calls. Detailed construction protocols and statistics are provided in Appendix \ref{sec: domaineval}, \ref{sec: domain dataset construction} and \ref{sec: Domain Stats}.

\subsection{Evaluation Framework}

\paragraph{Task Definition} Our evaluation follows best practices defined in HumanEval \citep{chen2021evaluating}, where models are required to implement a given function in each round. However, CodeFlowBench introduces distinct differences in the supporting materials provided to the models. We include the function signature \(F_t\), problem description \(S_t\) and pre-implemented functions for code reuse. Additionally, we provide an overarching background description of the problem as \(B\). This reflects the real-world scenario where developers often possess overall understanding of the entire problem when implementing code incrementally. As illustrated in Fig. \ref{codeflowbench}, models must implement target functions, \(C_t\) at turn \(t\), by leveraging pre-implemented components from prior turns \(\{C_1,...,C_{t-1}\}\). 

For baseline comparison, we define the single-turn variant where the model generates all components simultaneously. The mathematical expressions of both settings are presented below:
\begin{equation*}
\begin{aligned}
C_{1:T} & \leftarrow p\Bigl(\cdot \mid F_{1:T}, S_{1:T}, B\Bigr). \\
C_t & \leftarrow p\Bigl(\cdot \mid F_t, S_t, C_{<t}, B\Bigr), \quad 1 \leq t \leq T.
\end{aligned}
\end{equation*}


\paragraph{Performance Metrics}
We adopt widely-used Pass@k \citep{chen2021evaluating} as the main metric for both multi-turn and single-turn cases, but augment it with novel diagnostics for multi-turn analysis. While Pass@k evaluates final success rates, the coarse-grained nature fails to capture partial progress in failed attempts -two models may fail solve the entire problem at different stages but receive identical scores.

To address this limitation, we propose a new metric, Pass Turn ($PT$), which identifies the exact turn at which a model fails by leveraging unit tests for each subproblem. However, only considering failing turns may be biased due to the arbitrary topological ordering at the same depth. We further define, Pass Depth ($PD = D-d$), for bottom-up programming, where $d$ and $D$ are working depth and total depth. For statistical significance, we define the Average Pass Turn (APT) and Average Pass Depth (APD) by averaging across problems grouped by turn or depth. For multiple trials, we define APD@k and APT@k following the Pass@k expression. These metrics clearly distinguishing a model's strong performance on shallow steps from its potential struggles with deep dependencies. Please refer to Appendix \ref{sec: mathematical expression} for more details.




\section{Experiments}
\label{sec: Experiments}

\begin{table*}[t]
\centering

\scriptsize

\begin{tabular*}{\textwidth}{@{\extracolsep{\fill}}l c c c c c c c}
    
\toprule[1.5pt]
\multirow{2}{*}{\textbf{Model}} & 
\multicolumn{2}{c}{\textbf{Pass@1 (\%)}} & 
\multicolumn{5}{c}{\textbf{APD@1 (Average Pass Depth)}} \\
\cmidrule(lr){2-3} \cmidrule(lr){4-8}
& \textbf{Multi-Turn} & \textbf{Single-Turn} & \textbf{Overall} & \textbf{Depth1} & \textbf{Depth2} & \textbf{Depth3} & \textbf{Depth4} \\

\midrule
\multicolumn{8}{c}{\textbf{\textsc{CodeFlowBench-Comp}}} \\
\midrule

\rowcolor{sectiongray} \multicolumn{8}{l}{\textbf{\textit{Closed-Source}}} \\
GPT-5                      & 26.5 & 37.6 & 0.762 & 0.185 & 0.866 & 1.306 & 1.542 \\
GPT-5-mini                 & 19.1 & 38.0 & 0.655 & 0.103 & 0.748 & 1.200 & 1.375 \\
o3-mini                    & 22.7 & 38.9 & 0.570 & 0.322 & 0.585 & 0.818 & 1.250 \\
o1-mini                    & 20.8 & 37.8 & 0.541 & 0.233 & 0.581 & 0.818 & 1.125 \\
GPT-4.1-mini               & 24.4 & 38.7 & 0.602 & 0.265 & 0.673 & 0.873 & 1.042 \\
GPT-4o-mini                & 13.8 & 22.0 & 0.423 & 0.138 & 0.438 & 0.697 & 1.167 \\
Gemini-2.5-flash-thinking  & 22.7 & 39.8 & 0.642 & 0.167 & 0.761 & 0.981 & 1.208 \\
Gemini-3-flash-thinking    & \textbf{48.4} & \textbf{65.5} & \textbf{1.120} & \textbf{0.445} & \textbf{1.303} & \textbf{1.625} & \textbf{1.625} \\
Claude-4-Sonnet            & 27.3 & 42.9 & 0.830 & 0.185 & 0.893 & 1.250 & 1.375 \\
Claude-4.5-Sonnet          & 31.6 & 47.9    & 0.857 & 0.222 & 1.008 & 1.377 & 1.521 \\

\rowcolor{sectiongray} \multicolumn{8}{l}{\textbf{\textit{Open-Source (7B)}}} \\
Llama-3.1-8B-Instruct      & 0.9 & 3.5 & 0.208 & 0.011 & 0.224 & 0.412 & 0.792 \\
Qwen2.5-Coder-7B-Instruct  & 2.3 & 15.0 & 0.233 & 0.018 & 0.247 & 0.436 & 0.750 \\
Qwen3-8B                   & \textbf{9.1} & \textbf{18.3} & \textbf{0.406} & \textbf{0.088} & \textbf{0.444} & \textbf{0.738} & \textbf{1.125} \\

\rowcolor{sectiongray} \multicolumn{8}{l}{\textbf{\textit{Open-Source (32B)}}} \\
Qwen2.5-Coder-32B-Instruct & 8.6 & 19.8 & 0.316 & 0.067 & 0.342 & 0.570 & \textbf{0.917} \\
Qwen3-Coder-30B-A3B        & \textbf{11.6} & \textbf{26.0} & \textbf{0.405} & \textbf{0.100} & \textbf{0.458} & \textbf{0.681} & \textbf{0.917} \\

\rowcolor{sectiongray} \multicolumn{8}{l}{\textbf{\textit{Open-Source (70B)}}} \\
Llama-3.3-70B-Instruct     & \textbf{15.0} & \textbf{27.6} & \textbf{0.448} & \textbf{0.163} & \textbf{0.465} & \textbf{0.733} & \textbf{1.042} \\
Qwen2.5-72B-Instruct       & 9.1 & 21.3 & 0.301 & 0.110 & 0.314 & 0.497 & 0.583 \\

\rowcolor{sectiongray} \multicolumn{8}{l}{\textbf{\textit{Open-Source (Large)}}} \\
Deepseek-V3                & 18.0 & 35.7 & 0.529 & 0.219 & 0.549 & 0.836 & \textbf{1.208} \\
Deepseek-R1                & \textbf{20.5} & \textbf{46.1} & \textbf{0.569} & \textbf{0.303} & \textbf{0.606} & \textbf{0.842} & 0.916 \\

\midrule
\multicolumn{8}{c}{\textbf{\textsc{CodeFlowBench-Repo}}} \\
\midrule

\rowcolor{sectiongray} \multicolumn{8}{l}{\textbf{\textit{Closed-Source}}} \\
GPT-5                      & \textbf{34.7} & 48.8 & 0.898 & 0.483 & 1.020 & 1.385 & 3.200 \\
GPT-5-mini                 & 28.1 & 53.5 & 0.740 & 0.362 & 0.725 & 1.615 & 3.000 \\
o3-mini                    & 28.3 & 51.2 & 0.850 & 0.466 & 0.902 & 1.615 & 2.800 \\
o1-mini                    & 21.4 & 48.8 & 0.817 & 0.362 & 0.860 & 1.615 & \textbf{3.600} \\
GPT-4.1-mini               & 27.6 & 48.0 & 0.724 & 0.431 & 0.705 & \textbf{1.692} & 1.800 \\
GPT-4o-mini                & 24.6 & 43.3 & 0.667 & 0.414 & 0.800 & 1.000 & 1.400 \\
Gemini-2.5-flash-thinking  & 26.1 & 44.6 & 0.765 & 0.375 & 0.833 & 1.167 & \textbf{3.600} \\
Gemini-3-flash-thinking    & 33.9 & 56.7 & 0.905 & \textbf{0.500} & 1.020 & 1.308 & 3.400 \\
Claude-4-Sonnet            & 33.9 & \textbf{57.6} & 0.850 & 0.461 & 0.980 & 1.385 & 2.600 \\
Claude-4.5-Sonnet          & 33.1 & 56.7 & \textbf{0.913} & 0.448 & \textbf{1.039} & 1.615 & 3.200 \\

\rowcolor{sectiongray} \multicolumn{8}{l}{\textbf{\textit{Open-Source (7B)}}} \\
Llama-3.1-8B-Instruct      & 17.3 & 39.6 & 0.417 & \textbf{0.310} & 0.451 & 0.615 & 0.800 \\
Qwen2.5-Coder-7B-Instruct  & \textbf{19.7} & \textbf{39.1} & \textbf{0.528} & \textbf{0.310} & \textbf{0.647} & \textbf{0.692} & \textbf{1.400} \\
Qwen3-8B                   & 18.1 & 36.2 & 0.394 & 0.293 & 0.450 & 0.462 & 0.800 \\

\rowcolor{sectiongray} \multicolumn{8}{l}{\textbf{\textit{Open-Source (32B)}}} \\
Qwen2.5-Coder-32B-Instruct & \textbf{25.2} & \textbf{45.7} & \textbf{0.646} & 0.379 & \textbf{0.784} & 1.000 & \textbf{1.400} \\
Qwen3-Coder-30B-A3B        & 23.6 & 42.5 & 0.543 & \textbf{0.414} & 0.470 & \textbf{1.385} & 0.600 \\

\rowcolor{sectiongray} \multicolumn{8}{l}{\textbf{\textit{Open-Source (70B)}}} \\
Llama-3.3-70B-Instruct     & 22.1 & 39.6 & 0.614 & 0.345 & 0.522 & \textbf{1.846} & 1.400 \\
Qwen2.5-72B-Instruct       & \textbf{25.2} & \textbf{43.3} & \textbf{0.716} & \textbf{0.397} & \textbf{0.745} & 1.230 & \textbf{2.800} \\

\rowcolor{sectiongray} \multicolumn{8}{l}{\textbf{\textit{Open-Source (Large)}}} \\
Deepseek-V3                & 29.1 & 45.1 & \textbf{0.771} & \textbf{0.483} & 0.745 & \textbf{1.538} & 2.400 \\
Deepseek-R1                & \textbf{29.4} & \textbf{52.0} & 0.770 & 0.431 & \textbf{0.760} & \textbf{1.538} & \textbf{2.800} \\
\bottomrule[1.5pt]
\end{tabular*}
\caption{Performance on CodeFlowBench-Comp and CodeFlowBench-Repo across different model scales. Both datasets report Pass@1 and APD@1 (Average Pass Depth) across varying depths.}
\label{tab:model_comparison}
\end{table*}

\paragraph{Experiment Setup} 
For comprehensive experiments, we evaluate both close-sourced and open-sourced models. The closed‑source models include the GPT \citep{gpt5}, Gemini \citep{gemini3} and Claude families \citep{claude4}. The open‑source models include the Qwen \citep{yang2025qwen3}, DeepSeek \citep{liu2024deepseek,guo2025deepseek} and Llama families \citep{grattafiori2024llama}. We experiments on \textit{CodeFlowBench-Comp} and \textit{CodeFlowBench-Repo} under identical settings. For \textit{CodeFlowBench-Comp}, we selected a test subset comprising the most recent 1,000 problems to eliminate potential data leakage and decrease evaluation overhead while maintaining statistical significance. All models are evaluated in both multi-turn and single-turn scenarios, utilizing Pass@k and APD@k as the primary metrics. Please refer to Appendix \ref{sec: Experiment Setting} and \ref{sec:Instruction Templates} for implementation details and inference hyperparameter.

\subsection{Main Experiments}

Table \ref{tab:model_comparison} presents the main experiment results. The statistic rigor of evaluation result is reported in Appendix \ref{sec:Statistic Rigor}. For the majority of LLMs, the Pass@1 in the multi-turn scenario remains below 30\%, underscoring CodeFlowBench's difficulty. The performance across CodeFlowBench also reveals divergent model specializations. For example, Gemini-3-flash performs best on CodeFlowBench-Comp while GPT-5 outperforms on CodeFlowBench-Repo. Furthermore, we present more analysis as follows.


\paragraph{Multi-Turn versus Single-Turn} Table \ref{tab:model_comparison} reveals a substantial performance gap between multi-turn and single-turn scenarios across both \textit{CodeFlowBench-Comp} and \textit{CodeFlowBench-Repo}. This consistent degradation underscores the inherent complexity of multi-turn generation, which necessitates long-context coherence and robust dependency management. Notably, we observe a divergence in model behaviors: while reasoning-enhanced models like Deepseek-R1 and o3-mini achieve impressive single-turn scores, they suffer performance drops exceeding 50\% in multi-turn settings. In contrast, models such as GPT-5 demonstrate greater stability with narrower gaps. This distinction suggests that single-turn metrics largely reflect peak reasoning capabilities, whereas multi-turn evaluation exposes previously under-explored limitations in iterative development and dependency management.

\begin{table*}[t]
\centering

\scriptsize
\begin{tabular*}{\textwidth}{@{\extracolsep{\fill}}l ccccc cccc}
\toprule[1pt]
\multirow{2}{*}[-0.3em]{\textbf{Model}} & 
\multicolumn{5}{c}{\textbf{\textsc{CodeFlowBench-Comp}}} & 
\multicolumn{4}{c}{\textbf{\textsc{CodeFlowBench-Repo}}} \\
\cmidrule(lr){2-6} \cmidrule(lr){7-10}
 & \textbf{Overall} & \textbf{Turn 1} & \textbf{Turn 2} & \textbf{Turn 3} & \textbf{Turn 4} & 
\textbf{Overall} & \textbf{Turn 1} & \textbf{Turn 2} & \textbf{Turn 3+} \\
\midrule

\rowcolor{sectiongray} \multicolumn{10}{l}{\textbf{\textit{Closed-Source}}} \\
GPT-5                & 0.832 & 0.185 & 0.951 & 1.224 & 1.781 & 0.890 & \textbf{0.509} & 1.049 & 1.000 \\
GPT-5-mini                & 0.717 & 0.103 & 0.832 & 1.158 & 1.345 & 0.520 & 0.363 & 0.537 & 0.667 \\
o3-mini                   & 0.600 & 0.322 & 0.632 & 0.777 & 0.983 & 0.780 & 0.455 & 0.854 & 1.056 \\
o1-mini                   & 0.581 & 0.233 & 0.645 & 0.798 & 0.879 & 0.675 & 0.345 & 0.725 & 0.944 \\
GPT-4.1-mini              & 0.646 & 0.265 & 0.760 & 0.803 & 1.034 & 0.701 & 0.436 & 0.732 & 1.056 \\
GPT-4o-mini               & 0.467 & 0.137 & 0.501 & 0.638 & 0.982 & 0.571 & 0.418 & 0.650 & 0.778 \\
Gemini-2.5-flash-thinking & 0.695 & 0.167 & 0.855 & 0.923 & 1.218 & 0.687 & 0.364 & 0.879 & 0.600 \\
Gemini-3-flash-thinking   & \textbf{1.178} & \textbf{0.445} & \textbf{1.386} & \textbf{1.617} & \textbf{1.800} & 0.905 & 0.491 & 1.000 & 1.000 \\
Claude-4-Sonnet           & 0.830 & 0.185 & 0.990 & 1.153 & 1.564 & 0.866 & 0.455 & 1.024 & \textbf{1.389} \\
Claude-4.5-Sonnet         & 0.927 & 0.222 & 1.104 & 1.291 & 1.704 & \textbf{0.961} & 0.418 & \textbf{1.146} & 1.167 \\

\rowcolor{sectiongray} \multicolumn{10}{l}{\textbf{\textit{Open-Source (7B)}}} \\
Llama-3.1-8B-Instruct     & 0.232 & 0.011 & 0.245 & 0.404 & 0.534 & 0.370 & \textbf{0.327} & 0.439 & 0.389 \\
Qwen2.5-Coder-7B-Instruct & 0.258 & 0.018 & 0.270 & 0.394 & 0.638 & \textbf{0.528} & \textbf{0.327} & \textbf{0.512} & \textbf{0.778} \\
Qwen3-8B                  & \textbf{0.452} & \textbf{0.075} & \textbf{0.490} & \textbf{0.692} & \textbf{1.036} & 0.362 & 0.309 & 0.488 & 0.333 \\

\rowcolor{sectiongray} \multicolumn{10}{l}{\textbf{\textit{Open-Source (32B)}}} \\
Qwen2.5-Coder-32B-Instruct & 0.352 & 0.067 & 0.391 & 0.532 & 0.569 & \textbf{0.583} & 0.364 & \textbf{0.780} & \textbf{0.833} \\
Qwen3-Coder-30B-A3B        & \textbf{0.449} & \textbf{0.100} & \textbf{0.508} & \textbf{0.699} & \textbf{0.836} & 0.496 & \textbf{0.418} & 0.561 & 0.333 \\

\rowcolor{sectiongray} \multicolumn{10}{l}{\textbf{\textit{Open-Source (70B)}}} \\
Llama-3.3-70B-Instruct     & \textbf{0.493} & \textbf{0.163} & \textbf{0.515} & \textbf{0.681} & \textbf{1.000} & 0.516 & 0.345 & 0.553 & 0.750 \\
Qwen2.5-72B-Instruct       & 0.330 & 0.110 & 0.350 & 0.452 & 0.517 & \textbf{0.669} & \textbf{0.364} & \textbf{0.780} & \textbf{0.889} \\

\rowcolor{sectiongray} \multicolumn{10}{l}{\textbf{\textit{Open-Source (Large)}}} \\
Deepseek-V3                & 0.572 & 0.219 & 0.622 & 0.750 & \textbf{0.966} & \textbf{0.709} & \textbf{0.455} & 0.829 & \textbf{0.722} \\
Deepseek-R1                & \textbf{0.609} & \textbf{0.304} & \textbf{0.677} & \textbf{0.766} & \textbf{0.966} & 0.675 & 0.418 & \textbf{0.925} & 0.611 \\

\bottomrule[1pt]
\end{tabular*}
\caption{Pass Turn comparison across models on CodeFlowBench-Comp and CodeFlowBench-Repo. Consistent with Table \ref{tab:model_comparison}, the overall performance distribution shows that models with high APD@1 scores also exhibit high APT@1 scores, validating the precision of our metric.}

\label{tab:multi_turn_pass_comparison}
\end{table*}

\paragraph{Depth-Wise Performance} Our analysis across dependency depths reveals a dichotomy in scaling laws. First, in shallow dependencies, reasoning paradigms supersede parameter scale. Efficient reasoning models, such as o3-mini, significantly outperform larger base models like GPT-4o on \textit{CodeFlowBench-Comp}, suggesting that logical deduction is more critical than raw capacity for immediate or localized tasks. Second, deep dependencies function as a \textbf{capacity filter}, where model scale becomes decisive. In \textit{CodeFlowBench-Repo}, Qwen2.5-72B-Instruct doubles the performance of its 7B and 32B counterparts, indicating that maintaining coherence across extended dependency chains requires substantial parameter capacity. Finally, SOTA models like Gemini-3-flash-thinking and GPT-5 demonstrate full-spectrum dominance, eliminating trade-offs by combining agile reasoning for local logic with the extensive capacity required for global integration.From another perspective, we report APT@1 in Table \ref{tab:multi_turn_pass_comparison}. Notably, the APT trends closely align with APD, reinforcing our observations of models' depth-wise performance.

\sizhe{
Furthermore, we conduct extensive ablation studies to validate our evaluation framework. Specifically, we analyze: (1) the impact of the multi-turn paradigm versus monolithic generation; (2) the effectiveness of our data contamination mitigation; (3) the effect of the ``fail-stop'' mechanism; and (4) model sensitivity to different topological orderings of subproblems. Detailed results are provided in Appendix \ref{sec: Ablation Study}.
}


\subsection{Analysis and Discussion}
\begin{figure*}
    \centering
    \includegraphics[width=0.9\linewidth]{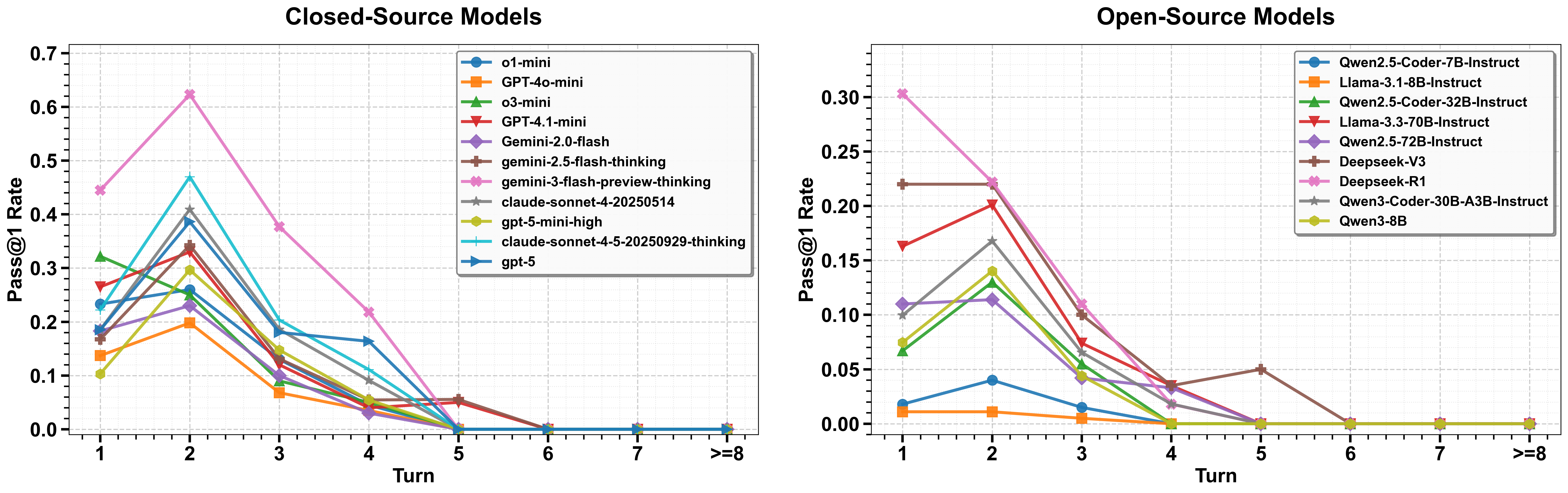}
    \caption{Models' Pass@1 results on multi-turn problems grouped by model categories and turn number.}
    \label{fig:Multi-turn Pass@1}
\end{figure*}

\paragraph{Dependency Structure Challenges in Multi-turn Scenarios} A deeper analysis of solved problems reveals a striking imbalance: the majority of correctly addressed cases correspond to problems with simple, linear dependency structures (e.g., shallow call graphs or sequential compositions). 
However, as problem architectures evolve toward modular and hierarchical dependencies (e.g., nested function calls, interdependent components), models exhibit significant performance degradation. This phenomenon is empirically validated in Figure \ref{fig:Multi-turn Pass@1}, which illustrates the Pass@1 scores across varying turn counts. The consistent performance trajectory demonstrates the inherent challenges in multi-turn code generation, where curves initially high for 1-2 turn problems and followed by sharp decline as turn counts increase.
Even top-performing models fail to solve problems requiring more than six turns. 
This underscores the critical limitation to balance local correctness and global integration across iterative development cycles.

\begin{figure}[htbp]
    \centering
    \includegraphics[width=\linewidth]{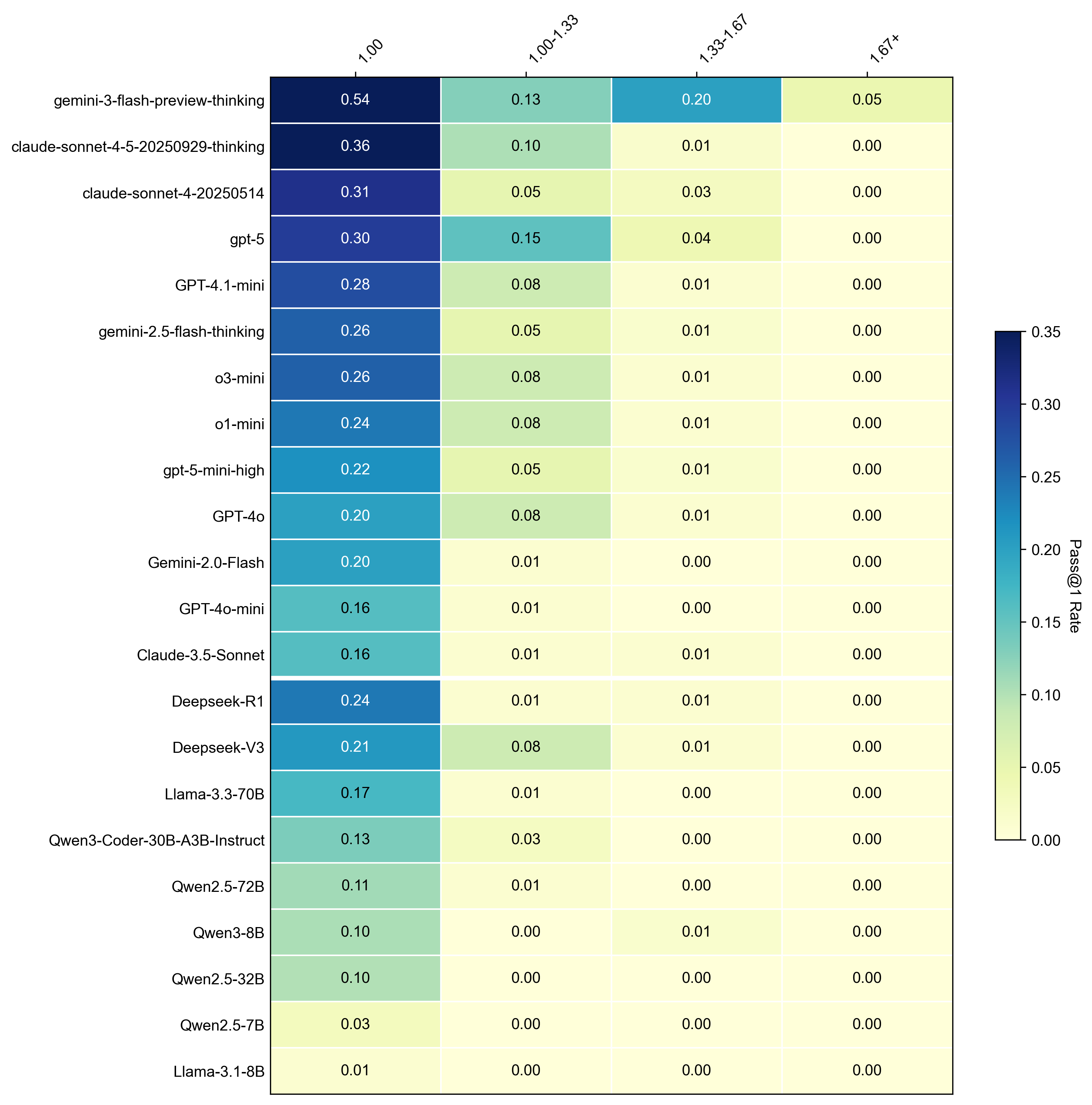
    }
    \caption{Heatmap of models' pass@1 scores on multi-turn problems within different DSC intervals.}
    \label{fig:DSC-pass1}
\end{figure}

To quantity dependency structure complexity, we introduce the \textit{Dependency Structure Complexity (DSC)} metric, defined as the ratio of total turns to the maximum depth in the AST. Figure \ref{fig:DSC-pass1} presents the models' performance across different DSC intervals, revealing that most models perform well on problems with linear dependency structures but struggle significantly as the dependency structure becomes more complex. 
For a problem, the DSC metric is defined as:
\[
\text{DSC(problem)}=\frac{\text{Overall-Turns(problem)}}{\text{Overall-Depth(problem)}}
\]
Recall that the overall-turn and overall-depth of a problem are derived from its AST, corresponding to the number of nodes and the depth of its AST. Based on this, we can see that a high \textit{DSC} value indicates a problem with a complex dependency structure. Figure \ref{fig:DSC-pass1} presents the \textit{pass@1} scores of models across different \textit{DSC} intervals. It can be observed that most models are only capable of solving problems with \textit{DSC} equal to 1, which corresponds to a simple linear dependency structure. Only a few leading models are able to solve a limited number of problems with \textit{DSC} values below 1.33. All models struggle significantly when faced with problems involving more complex structures. 

\paragraph{Fine-Grained Error Types in Multi-turn Generation}
Given the significant performance gap between models in multi-turn and single-turn scenarios, we conducted studies to identify the underlying reasons. We categorized errors into three primary types:
(1) \textbf{Incomplete Reasoning (IR):} Models often handle only straightforward "happy-path" cases and fail to generalize. They may oversimplify key requirements, omit boundary or atypical cases, or choose naive algorithms whose logic or performance collapses on larger inputs. This reflects a limitation in the models' reasoning abilities.
(2) \textbf{Insufficient Globalization (IG):} While a function's logic may run correctly in isolation, it may omit necessary imports, global constants, or shared-state interactions, preventing proper integration into the broader application or runtime. This indicates a limitation in the models' ability to manage global context.
(3) \textbf{Instruction Misinterpretation (IM):} Given multi-turn prompts, models could solve isolated subproblems but lack a coherent understanding of the overarching goal. Typical failures include misusing helper functions or implementing disorganized code within the top-level function, i.e. incorrect integration of components.
To quantify the distribution of errors, we randomly sampled examples from \textit{CodeFlowBench-Comp} and \textit{CodeFlowBench-Repo}. The error categories are manually annotated by authors. The proportion is calculated in Table \ref{tab:error-distribution}.

Table \ref{tab:error-distribution} offers several insights. First, the distinct error distributions confirm that our benchmark components target divergent primary challenges: \textit{CodeFlowBench-Comp} is dominated by IR and \textit{CodeFlowBench-Repo} emphasizes global context which shifting the burden to IG and IM. Second, distinct bottlenecks constrain models of different tiers. 
DeepSeek-V3 and o1-mini are primarily limited by IM, suggesting their performance is capped at instruction comprehension and adherence. 
In contrast, for SOTA models like Gemini-3-Flash, the bottleneck shifts back to IR, indicating task understanding becomes minor issues. 
Detailed case studies are provided in Appendix \ref{sec: Error Case Study}.

\begin{table}

\resizebox{\columnwidth}{!}{
\begin{tabular}{llcccc}
\toprule
Model & Dataset & IR(\%) & IG(\%) & IM(\%) & Others(\%) \\
\midrule
\multirow{2}{*}{DeepSeek-V3} & Comp & 47.2 & 14.6 & 32.6 & 5.6 \\
                             & Repo & 32.1 & 22.6 & 44.0 & 1.2 \\
\midrule
\multirow{2}{*}{o1-mini}     & Comp & 38.7 & 11.8 & 43.0 & 6.5 \\
                             & Repo & 23.7 & 18.3 & 55.9 & 2.2 \\
\midrule
\multirow{2}{*}{Gemini-3-Flash} & Comp & 52.3 & 20.5 & 27.3 & 0.0 \\
                                & Repo & 34.6 & 21.8 & 41.0 & 2.6 \\
\bottomrule
\end{tabular}
}
\caption{Error distribution across models and benchmarks. Values indicate the percentage of each error category relative to the total number of errors identified in the sampled dataset for each model.}
\label{tab:error-distribution}
\end{table}

\section{Conclusion}
\label{sec: Conclusion}
This paper introduces CodeFlowBench, the first benchmark specifically designed to evaluate multi-turn iterative code generation capabilities in realistic development workflows, i.e., the \textit{codeflow}. 
CodeFlowBench consists of two subsets: CodeFlowBench-Comp for multi-step algorithmic reasoning via competition challenges, and CodeFlowBench-Repo for evaluating generalization in real-world software contexts.
Our benchmark makes three key contributions: (1) an automated pipeline for decomposing complex problems into dependency-aware subproblems with paired unit tests, (2) a novel evaluation framework with proposed structural metrics, such as APD@k and DSC, to quantify multi-turn performance, and (3) the discovery of substantial performance gaps between multi-turn and single-turn scenarios (up to 50\% performance degradation). Our fine-grained analysis identifies dominant failure modes and provides insights for further advancements. Extensive experiments across 19 popular LLMs highlight the substantial challenges posed by both the codeflow task and our benchmark. We believe CodeFlowBench not only illuminates critical limitations in existing LLMs but also paves the way for more realistic and powerful code generation systems.

\clearpage

\section*{Limitations}
We acknowledge two limitations that present opportunities for future expansion. First, while the pipeline's subproblem generation phase is general, the data preparation phase is currently specific. To achieve universal compatibility, we can consider implementing an agentic framework for automatically adapting diverse codebases. This will expand our pipeline's coverage to a broader range of data sources beyond Codeforces and GitHub, such as Bitbucket, establishing a truly source-agnostic, end-to-end framework. Second, we currently lack models specifically fine-tuned for the codeflow task. Future work will consider fine-tuning specialized models and agentic systems on this data to assist developers in real-world, iterative programming scenarios. We also aim to extend our codeflow paradigm to include iterative debugging loops. By training specialized systems on these dynamic scenarios, we seek to enable robust and sustainable code implementation and maintenance, allowing models to effectively resolve requirements through multi-turn refinement.

\section*{Ethical Consideration}
Codeforces provides an official “Codeforces materials usage license”\footnote{\url{https://codeforces.com/blog/entry/967}}. We have taken great care to make our pipeline fully comply with all licensing requirements and community guidelines:

Firstly, all problems we use are officially published by Codeforces, and we provide clear attribution to Codeforces. Our usage is strictly non-commercial and academic, in accordance with the license’s permitted use cases. Our data pipeline strictly avoids scraping any user-submitted code, forum discussion content or third-party solutions. We only collect officially published materials, namely problem statements and editorials.

Secondly, officially published Codeforces materials have long been used for widely recognized code generation benchmarks such as CodeContest  \citep{li2022competition}, LiveCodeBench \citep{jain2024livecodebench} and CodeElo \citep{quan2025codeelo}. Our work follows this community tradition and with the utmost respect for Codeforces.

Besides, we will also make our code properly hosted and clearly state its intended use to ensure responsible data collection.

\section*{Acknowledgements}
This work is supported by Fundamental and Interdisciplinary Disciplines Breakthrough Plan of the Ministry of Education of China (JYB2025XDXM113), National Natural Science Foundation of China (92470121, 62402016), National Key R\&D Program of China (2024YFA1014003), Zhongguancun Academy (C20250204, C20250602), Beijing Major Science and Technology Project (Z251100008125043, Z251100008425023), and High-performance Computing Platform of Peking University.

\bibliography{custom}

@article{feitosa2020code,
  title={Code reuse in practice: Benefiting or harming technical debt},
  author={Feitosa, Daniel and Ampatzoglou, Apostolos and Gkortzis, Antonios and Bibi, Stamatia and Chatzigeorgiou, Alexander},
  journal={Journal of Systems and Software},
  volume={167},
  pages={110618},
  year={2020},
  publisher={Elsevier}
}

@article{hendrycks2021measuring,
  title={Measuring coding challenge competence with apps},
  author={Hendrycks, Dan and Basart, Steven and Kadavath, Saurav and Mazeika, Mantas and Arora, Akul and Guo, Ethan and Burns, Collin and Puranik, Samir and He, Horace and Song, Dawn and others},
  journal={arXiv preprint arXiv:2105.09938},
  year={2021}
}

@article{yang2023intercode,
  title={Intercode: Standardizing and benchmarking interactive coding with execution feedback},
  author={Yang, John and Prabhakar, Akshara and Narasimhan, Karthik and Yao, Shunyu},
  journal={Advances in Neural Information Processing Systems},
  volume={36},
  pages={23826--23854},
  year={2023}
}

@article{guo2024deepseek,
  title={DeepSeek-Coder: When the Large Language Model Meets Programming--The Rise of Code Intelligence},
  author={Guo, Daya and Zhu, Qihao and Yang, Dejian and Xie, Zhenda and Dong, Kai and Zhang, Wentao and Chen, Guanting and Bi, Xiao and Wu, Yu and Li, YK and others},
  journal={arXiv preprint arXiv:2401.14196},
  year={2024}
}

@article{luo2023wizardcoder,
  title={Wizardcoder: Empowering code large language models with evol-instruct},
  author={Luo, Ziyang and Xu, Can and Zhao, Pu and Sun, Qingfeng and Geng, Xiubo and Hu, Wenxiang and Tao, Chongyang and Ma, Jing and Lin, Qingwei and Jiang, Daxin},
  journal={arXiv preprint arXiv:2306.08568},
  year={2023}
}

@article{islam2024mapcoder,
  title={Mapcoder: Multi-agent code generation for competitive problem solving},
  author={Islam, Md Ashraful and Ali, Mohammed Eunus and Parvez, Md Rizwan},
  journal={arXiv preprint arXiv:2405.11403},
  year={2024}
}

@article{guo2025deepseek,
  title={Deepseek-r1: Incentivizing reasoning capability in llms via reinforcement learning},
  author={Guo, Daya and Yang, Dejian and Zhang, Haowei and Song, Junxiao and Zhang, Ruoyu and Xu, Runxin and Zhu, Qihao and Ma, Shirong and Wang, Peiyi and Bi, Xiao and others},
  journal={arXiv preprint arXiv:2501.12948},
  year={2025}
}

@article{jiang2024survey,
  title={A survey on large language models for code generation},
  author={Jiang, Juyong and Wang, Fan and Shen, Jiasi and Kim, Sungju and Kim, Sunghun},
  journal={arXiv preprint arXiv:2406.00515},
  year={2024}
}

@article{liu2024deepseek,
  title={Deepseek-v3 technical report},
  author={Liu, Aixin and Feng, Bei and Xue, Bing and Wang, Bingxuan and Wu, Bochao and Lu, Chengda and Zhao, Chenggang and Deng, Chengqi and Zhang, Chenyu and Ruan, Chong and others},
  journal={arXiv preprint arXiv:2412.19437},
  year={2024}
}

@article{zhuo2024bigcodebench,
  title={Bigcodebench: Benchmarking code generation with diverse function calls and complex instructions},
  author={Zhuo, Terry Yue and Vu, Minh Chien and Chim, Jenny and Hu, Han and Yu, Wenhao and Widyasari, Ratnadira and Yusuf, Imam Nur Bani and Zhan, Haolan and He, Junda and Paul, Indraneil and others},
  journal={arXiv preprint arXiv:2406.15877},
  year={2024}
}

@article{chen2021evaluating,
  title={Evaluating large language models trained on code},
  author={Chen, Mark and Tworek, Jerry and Jun, Heewoo and Yuan, Qiming and Pinto, Henrique Ponde De Oliveira and Kaplan, Jared and Edwards, Harri and Burda, Yuri and Joseph, Nicholas and Brockman, Greg and others},
  journal={arXiv preprint arXiv:2107.03374},
  year={2021}
}

@article{yu2024humaneval,
  title={HumanEval Pro and MBPP Pro: Evaluating Large Language Models on Self-invoking Code Generation},
  author={Yu, Zhaojian and Zhao, Yilun and Cohan, Arman and Zhang, Xiao-Ping},
  journal={arXiv preprint arXiv:2412.21199},
  year={2024}
}

@article{jimenez2023swe,
  title={Swe-bench: Can language models resolve real-world github issues?},
  author={Jimenez, Carlos E and Yang, John and Wettig, Alexander and Yao, Shunyu and Pei, Kexin and Press, Ofir and Narasimhan, Karthik},
  journal={arXiv preprint arXiv:2310.06770},
  year={2023}
}

@article{austin2021program,
  title={Program synthesis with large language models},
  author={Austin, Jacob and Odena, Augustus and Nye, Maxwell and Bosma, Maarten and Michalewski, Henryk and Dohan, David and Jiang, Ellen and Cai, Carrie and Terry, Michael and Le, Quoc and others},
  journal={arXiv preprint arXiv:2108.07732},
  year={2021}
}

@article{quan2025codeelo,
  title={CodeElo: Benchmarking Competition-level Code Generation of LLMs with Human-comparable Elo Ratings},
  author={Quan, Shanghaoran and Yang, Jiaxi and Yu, Bowen and Zheng, Bo and Liu, Dayiheng and Yang, An and Ren, Xuancheng and Gao, Bofei and Miao, Yibo and Feng, Yunlong and others},
  journal={arXiv preprint arXiv:2501.01257},
  year={2025}
}

@article{jain2024livecodebench,
  title={Livecodebench: Holistic and contamination free evaluation of large language models for code},
  author={Jain, Naman and Han, King and Gu, Alex and Li, Wen-Ding and Yan, Fanjia and Zhang, Tianjun and Wang, Sida and Solar-Lezama, Armando and Sen, Koushik and Stoica, Ion},
  journal={arXiv preprint arXiv:2403.07974},
  year={2024}
}

@article{nijkamp2022codegen,
  title={Codegen: An open large language model for code with multi-turn program synthesis},
  author={Nijkamp, Erik and Pang, Bo and Hayashi, Hiroaki and Tu, Lifu and Wang, Huan and Zhou, Yingbo and Savarese, Silvio and Xiong, Caiming},
  journal={arXiv preprint arXiv:2203.13474},
  year={2022}
}

@article{huang2023agentcoder,
  title={Agentcoder: Multi-agent-based code generation with iterative testing and optimisation},
  author={Huang, Dong and Zhang, Jie M and Luck, Michael and Bu, Qingwen and Qing, Yuhao and Cui, Heming},
  journal={arXiv preprint arXiv:2312.13010},
  year={2023}
}

@article{grattafiori2024llama,
  title={The llama 3 herd of models},
  author={Grattafiori, Aaron and Dubey, Abhimanyu and Jauhri, Abhinav and Pandey, Abhinav and Kadian, Abhishek and Al-Dahle, Ahmad and Letman, Aiesha and Mathur, Akhil and Schelten, Alan and Vaughan, Alex and others},
  journal={arXiv preprint arXiv:2407.21783},
  year={2024}
}

@article{li2022competition,
  title={Competition-level code generation with alphacode},
  author={Li, Yujia and Choi, David and Chung, Junyoung and Kushman, Nate and Schrittwieser, Julian and Leblond, R{\'e}mi and Eccles, Tom and Keeling, James and Gimeno, Felix and Dal Lago, Agustin and others},
  journal={Science},
  volume={378},
  number={6624},
  pages={1092--1097},
  year={2022},
  publisher={American Association for the Advancement of Science}
}

@article{wang2025maintaincoder,
  title={MaintainCoder: Maintainable Code Generation Under Dynamic Requirements},
  author={Wang, Zhengren and Ling, Rui and Wang, Chufan and Yu, Yongan and Li, Zhiyu and Xiong, Feiyu and Zhang, Wentao},
  journal={arXiv preprint arXiv:2503.24260},
  year={2025}
}

@article{jin2024llms,
  title={From llms to llm-based agents for software engineering: A survey of current, challenges and future},
  author={Jin, Haolin and Huang, Linghan and Cai, Haipeng and Yan, Jun and Li, Bo and Chen, Huaming},
  journal={arXiv preprint arXiv:2408.02479},
  year={2024}
}

@article{liu2024large,
  title={Large language model-based agents for software engineering: A survey},
  author={Liu, Junwei and Wang, Kaixin and Chen, Yixuan and Peng, Xin and Chen, Zhenpeng and Zhang, Lingming and Lou, Yiling},
  journal={arXiv preprint arXiv:2409.02977},
  year={2024}
}

@article{li2024devbench,
  title={Devbench: A comprehensive benchmark for software development},
  author={Li, Bowen and Wu, Wenhan and Tang, Ziwei and Shi, Lin and Yang, John and Li, Jinyang and Yao, Shunyu and Qian, Chen and Hui, Binyuan and Zhang, Qicheng and others},
  journal={CoRR},
  year={2024}
}

@article{cram2019agile,
  title={Agile development in practice: Lessons from the trenches},
  author={Cram, W Alec},
  journal={Information Systems Management},
  volume={36},
  number={1},
  pages={2--14},
  year={2019},
  publisher={Taylor \& Francis}
}

@book{larman2004agile,
  title={Agile and iterative development: a manager's guide},
  author={Larman, Craig},
  year={2004},
  publisher={Addison-Wesley Professional}
}

@article{haefliger2008code,
  title={Code reuse in open source software},
  author={Haefliger, Stefan and Von Krogh, Georg and Spaeth, Sebastian},
  journal={Management science},
  volume={54},
  number={1},
  pages={180--193},
  year={2008},
  publisher={INFORMS}
}

@article{abrahamsson2017agile,
  title={Agile software development methods: Review and analysis},
  author={Abrahamsson, Pekka and Salo, Outi and Ronkainen, Jussi and Warsta, Juhani},
  journal={arXiv preprint arXiv:1709.08439},
  year={2017}
}

@article{huang2024effibench,
  title={Effibench: Benchmarking the efficiency of automatically generated code},
  author={Huang, Dong and Qing, Yuhao and Shang, Weiyi and Cui, Heming and Zhang, Jie},
  journal={Advances in Neural Information Processing Systems},
  volume={37},
  pages={11506--11544},
  year={2024}
}

@article{wei2023magicoder,
  title={Magicoder: Empowering code generation with oss-instruct},
  author={Wei, Yuxiang and Wang, Zhe and Liu, Jiawei and Ding, Yifeng and Zhang, Lingming},
  journal={arXiv preprint arXiv:2312.02120},
  year={2023}
}

@article{gu2024cruxeval,
  title={Cruxeval: A benchmark for code reasoning, understanding and execution},
  author={Gu, Alex and Rozi{\`e}re, Baptiste and Leather, Hugh and Solar-Lezama, Armando and Synnaeve, Gabriel and Wang, Sida I},
  journal={arXiv preprint arXiv:2401.03065},
  year={2024}
}

@book{boduch2019react,
  title={React material-ui cookbook: build captivating user experiences using react and material-ui},
  author={Boduch, Adam},
  year={2019},
  publisher={Packt Publishing Ltd}
}

@inproceedings{zhu2025domaineval,
  title={Domaineval: An auto-constructed benchmark for multi-domain code generation},
  author={Zhu, Qiming and Cao, Jialun and Lu, Yaojie and Lin, Hongyu and Han, Xianpei and Sun, Le and Cheung, Shing-Chi},
  booktitle={Proceedings of the AAAI Conference on Artificial Intelligence},
  volume={39},
  number={24},
  pages={26148--26156},
  year={2025}
}

@techreport{gpt5,
  title       = {GPT-5 System Card},
  author      = {{OpenAI}}, 
  institution = {OpenAI},
  year        = {2025},
  url         = {https://cdn.openai.com/gpt-5-system-card.pdf},
  type        = {System Card}
}

@techreport{claude4,
  title       = {Claude-4 System Card},
  author      = {{Anthropic}},
  institution = {Anthropic},
  year        = {2025},
  url         = {https://www-cdn.anthropic.com/4263b940cabb546aa0e3283f35b686f4f3b2ff47.pdf},
  type        = {System Card}
}

@techreport{gemini3,
  title       = {Gemini-3-flash System Card},
  author      = {{Google}},
  institution = {Google},
  year        = {2025},
  url         = {https://storage.googleapis.com/deepmind-media/Model-Cards/Gemini-3-Flash-Model-Card.pdf},
  type        = {System Card}
}

@article{yang2025qwen3,
  title={Qwen3 technical report},
  author={Yang, An and Li, Anfeng and Yang, Baosong and Zhang, Beichen and Hui, Binyuan and Zheng, Bo and Yu, Bowen and Gao, Chang and Huang, Chengen and Lv, Chenxu and others},
  journal={arXiv preprint arXiv:2505.09388},
  year={2025}
}

\clearpage
\appendix
\section*{Appendix}
\startcontents[sections]
\printcontents[sections]{l}{1}{\setcounter{tocdepth}{3}}

\section{Data Curation Pipeline}
\subsection{Detail of Problem Scraping}
\label{sec: Detail of Problem Scraping}
\begin{figure*}
    \centering
    \includegraphics[width=0.7\linewidth]{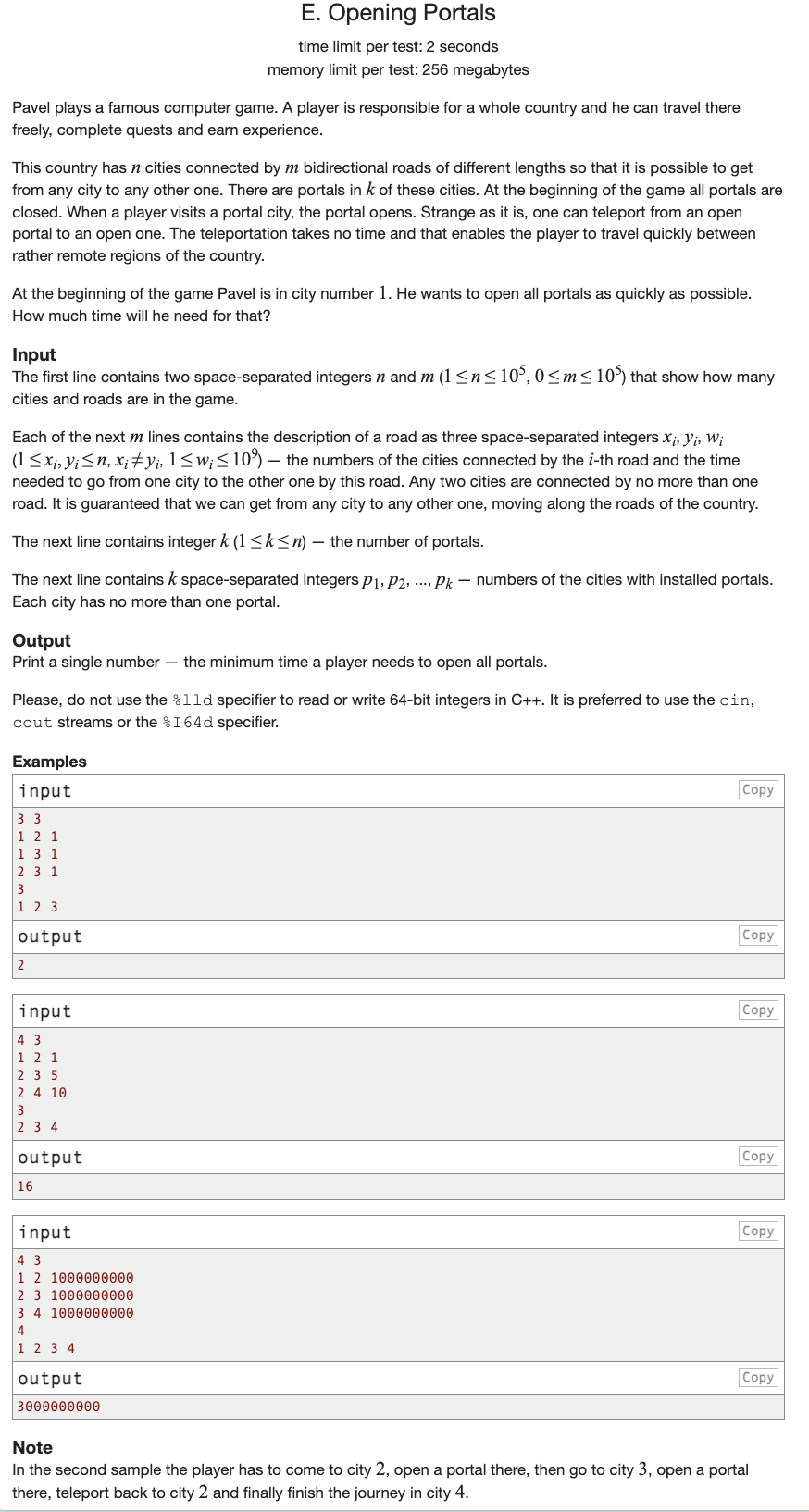}
    \caption{An example page of problems on Codeforces, which contains \textit{problem ID, title, time and memory complexity limits, problem description, input description, output description, sample tests, notes} of each problem.The original problem is \href{https://codeforces.com/problemset/problem/196/E}{196E}}.
    \label{fig:196E}
\end{figure*}

The original problem contain two parts. The first part is scraped from its corresponding codeforces official website. An example problem page is shown in Figure \ref{fig:196E}.

The second part is scraped by \textit{Problemsets.Problems}\footnote{\url{https://codeforces.com/api/problemset.problems}} API provided by Codeforces.We record the rating and tags of each problem. Rating is a metric that reflect the difficulty of the problem and tags is a list that contains knowledge scope of the problem.

\begin{figure*}
    \centering
    \includegraphics[width=1\linewidth]{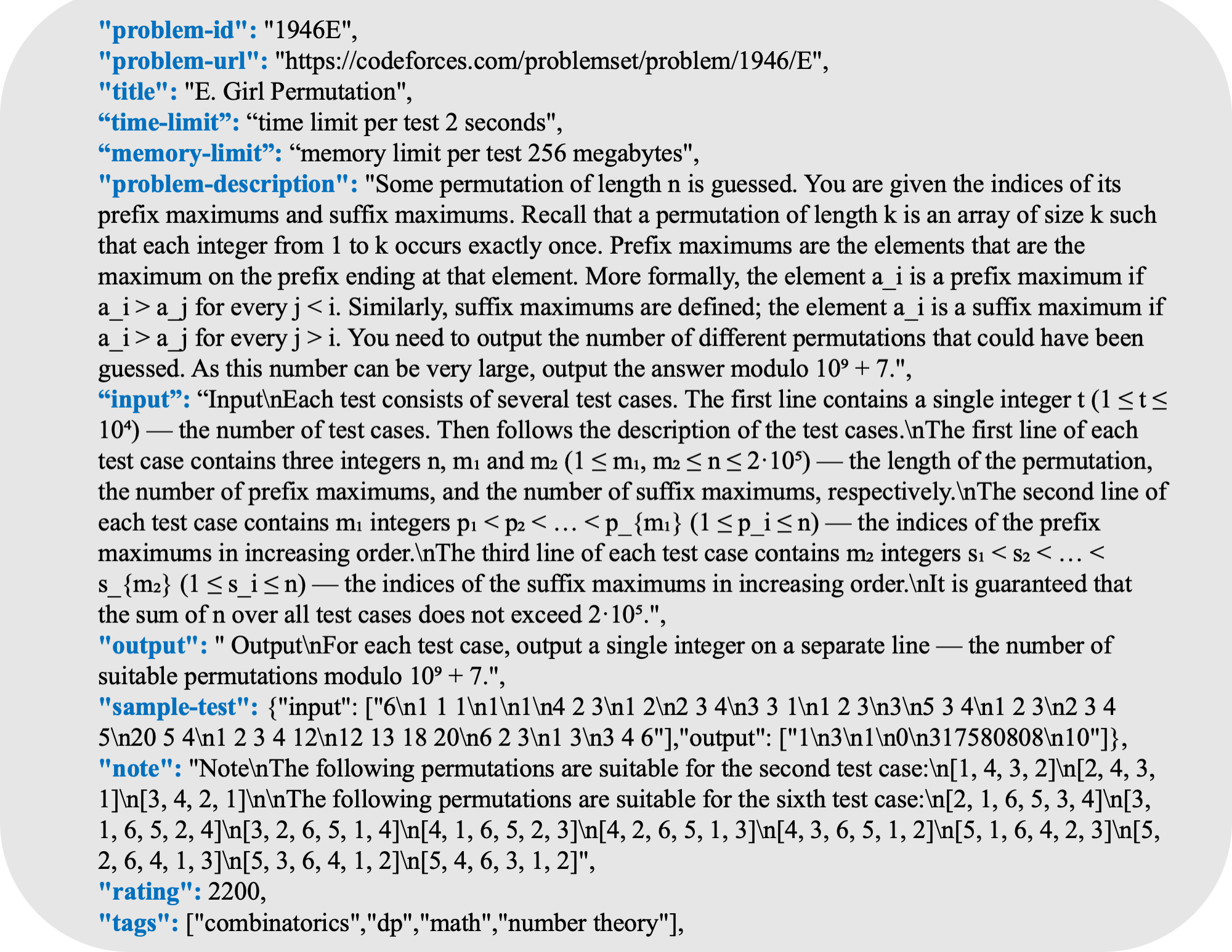}
    \caption{An example of original coding problem we obtained in stage I.To make the content more clear, we remove html denotions that contained in original content. The original problem is \href{https://codeforces.com/problemset/problem/1946/E}{1946E}, which is used for illustration in Figure \ref{pipeline}}.
    \label{fig:stage I demo}
\end{figure*}

Combine the above two part, we obtain a original coding problem for CodeFlowBench, a full example is shown in Figure \ref{fig:stage I demo}.
\subsection{Detail of Solution Scraping}
\label{sec: Detail of Solution Scraping}
During the problem‐scraping process, we also collected the links under the “Contest material” section on the right side of each problem page and identified which of those led to solution blogs. Crawling these editorial pages is fairly complex, because Codeforces’ official write‐ups are hosted as personal blogs whose formats can vary over time, making content extraction more challenging. Although each round’s problems (e.g. “123A,” “123B,” etc.) live on separate pages, all of the editorials for a given numeric ID usually appear on a single blog page. We therefore need to assign each sub‐problem’s write‐up (A, B, C, …) to its corresponding problem. To do this, our crawler first locates the distinct anchor points that mark each sub‐problem section, then extracts the content between each anchor and the next as that problem’s editorial. The benefit of this approach is that, while in the end we only need the solution code itself, editorials often consist of plain text explanations, a single code snippet, or multiple code variants and languages. Relying on a purely mechanical scraper makes it difficult to isolate exactly the code we want, so it’s more effective to pass the raw editorial content to an LLM for final organization and extraction.

In general, there are two ways to locate an anchor: by URL and by text. Blogs label problems in many different forms. However, most blogs make that label into a hyperlink pointing back to the original problem, which gives us a reliable way to identify the anchor.

\begin{figure*}[htbp]
  \centering
  \begin{subfigure}[t]{0.48\linewidth}
    \centering
    \includegraphics[width=\linewidth]{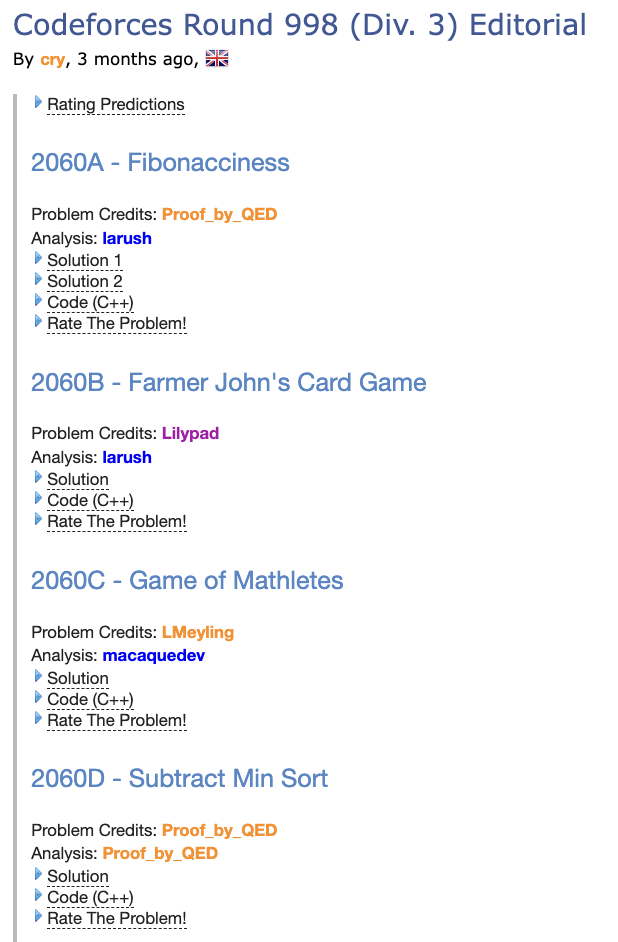}
    \caption{}
    \label{fig:stageII_url_demo}
  \end{subfigure}
  \hfill
  \begin{subfigure}[t]{0.45\linewidth}
    \centering
    \includegraphics[width=\linewidth]{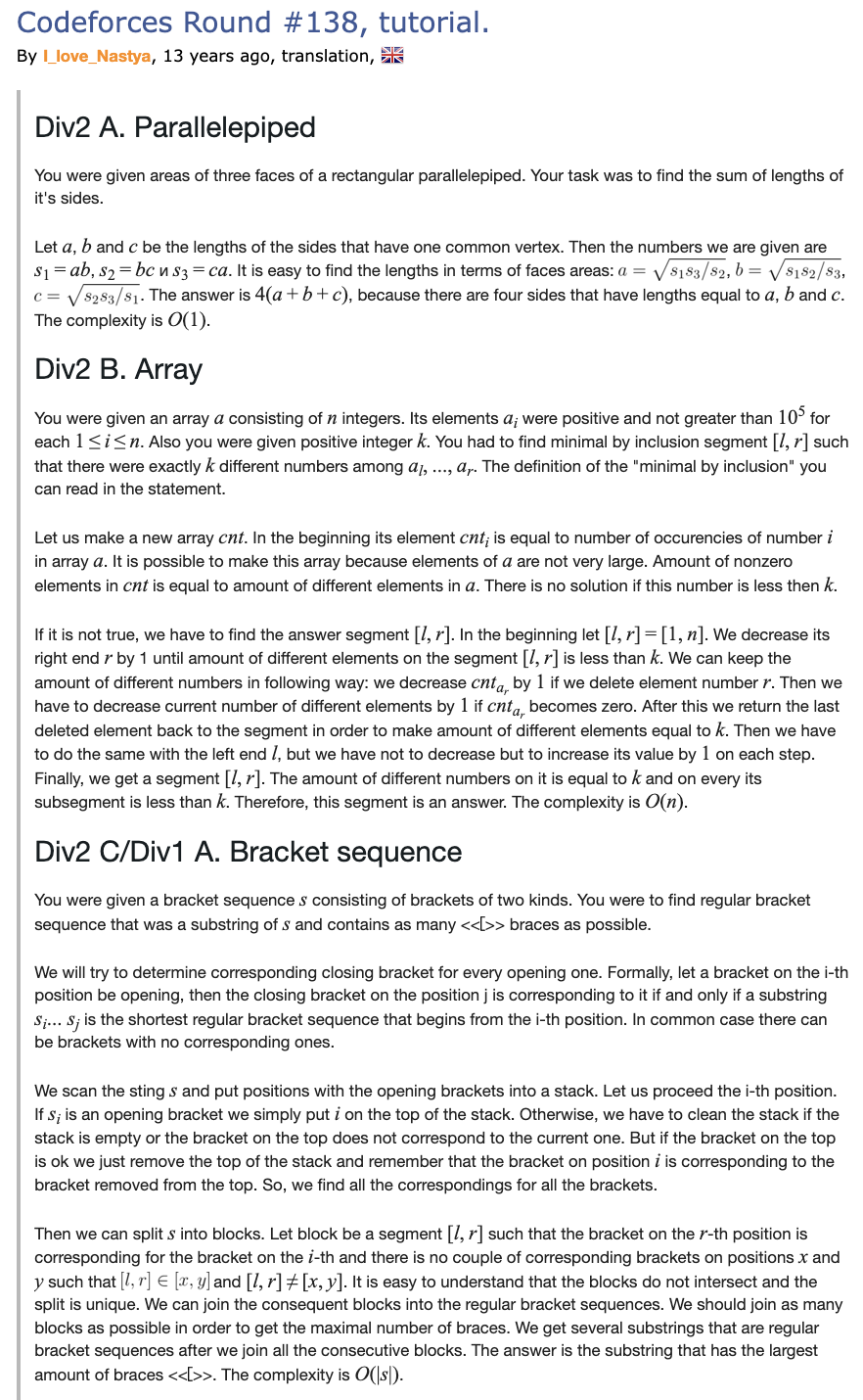}
    \caption{}
    \label{fig:stageII_div_demo}
  \end{subfigure}
  \caption{Subfigure (a) is an example of using URL anchor identification technique. The anchor here is each subtitle displayed as "2060A-Fibonacciness". The blue font color of such subtitle indicates a URL to original problem page is setted. In practice, the existence of URL is identified by analysing the HTML code of this website.For this kind of website, we identify these subtitles as anchors and scrape the content between each subtitle.Subfigure (b) is an example of using div anchor identification technique.The anchor here is each subtitle displayed as "Div2A. Parallelepiped".These kinds of subtitle occurs in a contest round that contain two div level problems.We use reference table to project the div notation to original problem id and then identify them as corresponding anchor.}
  \label{fig:stageII_anchor_demo}
\end{figure*}

Therefore, our primary and most precise method is to search for a URL containing the problem ID (for example, “problemset/problem/2060/A” or “contest/blog/2060/A”) and treat its position as the anchor. Once the anchor is found, we scan the surrounding page and extract its content to obtain the editorial.Anchor scanning and recording also relies on a problem‑ID reference table, which we built from the IDs of all problems scraped in the first step. Its main role is to guide the code when matching anchors: for instance, if the table shows that numeric ID 2060 has sub‑IDs “A” through “G”, the scraper first reads those sub‑IDs, then walks through the page using the URL‑based or text‑based method to record the exact anchor for each sub‑ID.An example that fit URL anchor identification technique in shown in Figure \ref{fig:stageII_url_demo}.

\begin{figure*}
    \centering
    \includegraphics[width=1\linewidth]{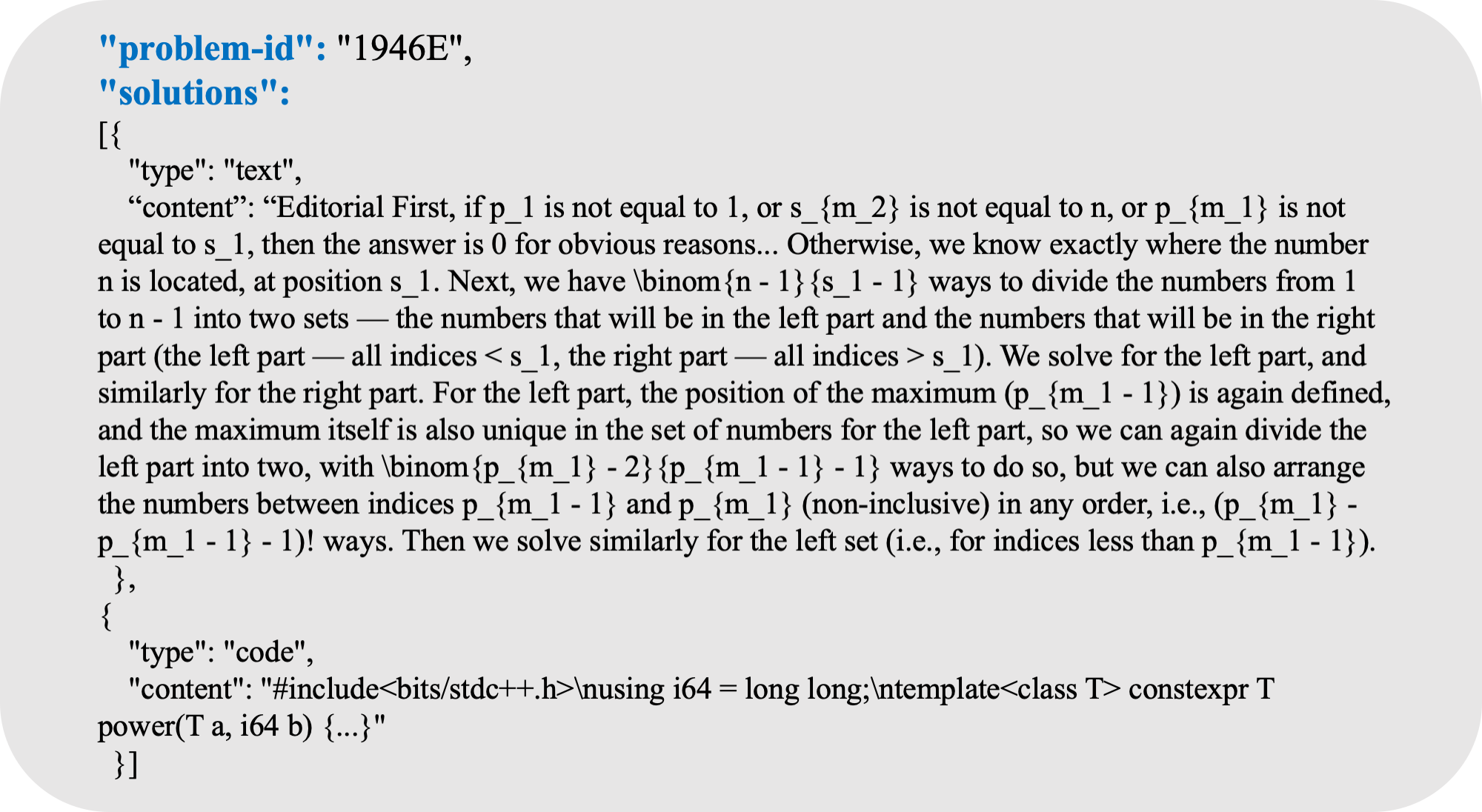}
    \caption{The scraped and processed solution we obtained in stage II. The original problem is \href{https://codeforces.com/problemset/problem/1946/E}{1946E}}.
    \label{fig:stage II demo}
\end{figure*}

However, there are still little parts of early editorial blogs didn’t include URL hyperlinks, so in those cases we fall back on regular‑expression–based text matching wherever possible.  Based on the formats we observed, we designed two main matching strategies: 
\begin{itemize}
    \item \textbf{Difficulty‑label matching.}  A number of blogs publish all of a round’s problems on one page and mark them with labels like “Div2” or “Div1” (since most rounds contain two Div2‑level problems and several Div1‑level ones).  To handle this, we use our problem‑ID reference table to identify all sub‑IDs belonging to the same numeric contest but with different difficulty levels, tag them accordingly in the table, and then, during the anchor‑matching process, if the scraper detects a difficulty label it will also try to match anchors based on that label. An example is shown in Figure \ref{fig:stageII_div_demo}.
    \item \textbf{Problem‑label matching.}  Beyond difficulty tags, many blogs use the literal “Problem A”, “Problem B”, etc. format.  We include a specific regex pattern to detect those “Problem+sub‑ID” labels and assign each section to the correct sub‑problem.
\end{itemize}
After crawling the editorial for each problem, we applied a series of processing steps to ensure quality.Firstly, we removed any editorials that were too short or empty—these problems were excluded from the dataset. To make it easier for an LLM to understand and process them, we then split each editorial into two parts: the code solution and the textual explanation, so that the model can consult the code first and then the accompanying text.The processed result is shown in Figure \ref{fig:stage II demo}.
\subsection{Detail of Solution Code Generation}
\label{sec: Detail of Solution Code Generation}
\begin{figure*}
    \centering
    \includegraphics[width=1\linewidth]{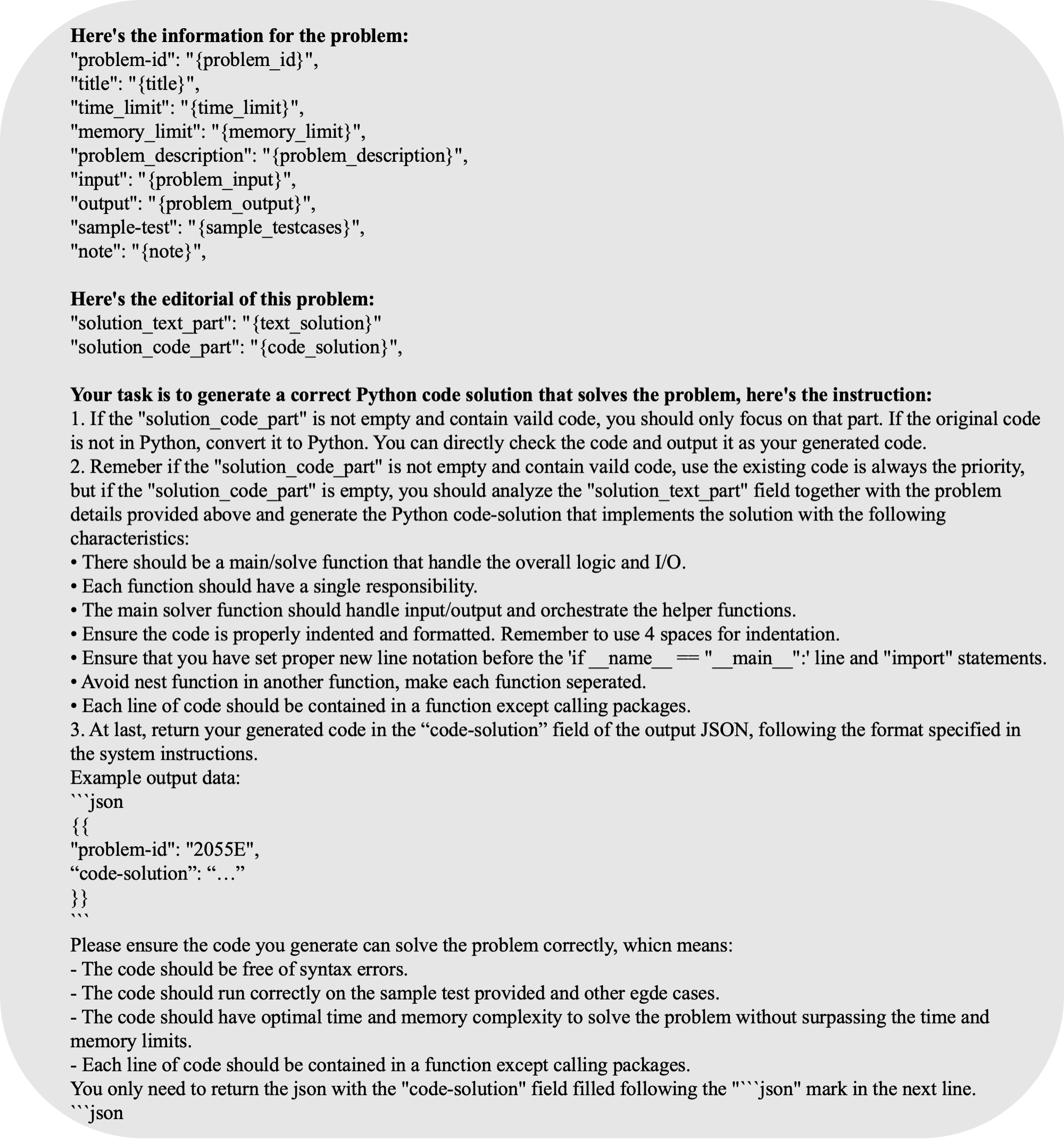}
    \caption{The prompt template used for code convertion in stage III.The whole content of the  example output data is not shown for length limitation.}
    \label{fig:Stage III prompt template}
\end{figure*}

Although we’ve already scraped the official solution for each problem, an LLM-based post-processing step is still required for two main reasons:\textbf{(1) Presence of “global code segments”.}
Some solutions include code that isn’t encapsulated in any function. We must split the entire codebase into multiple functions and ensure that the top-level function can fully solve the problem. Since these global segments can’t be recognized during standard parsing, we rely on an LLM to reorganize the provided code so that it becomes fully parsable.\textbf{(2)Early solutions exist only as text.}Some of the older official solutions consist solely of textual descriptions without any runnable code. We need an LLM to convert those narratives into executable code.The prompt template for code convert is shown in Figure \ref{fig:Stage III prompt template}
The Codeforces official judging system is used to verify code correctness. We employ an automated submission bot that navigates to the Codeforces submission page\footnote{\url{https://codeforces.com/problemset/submit}}, fills in all required fields, and submits the solution. The site will be automatically redirects to the results page after submission, from which we scrape the verification outcome.

\subsection{Detail of Subproblems Generation}
\label{sec: Detail of Subproblems Generation}
\begin{figure*}
    \centering
    \includegraphics[width=1\linewidth]{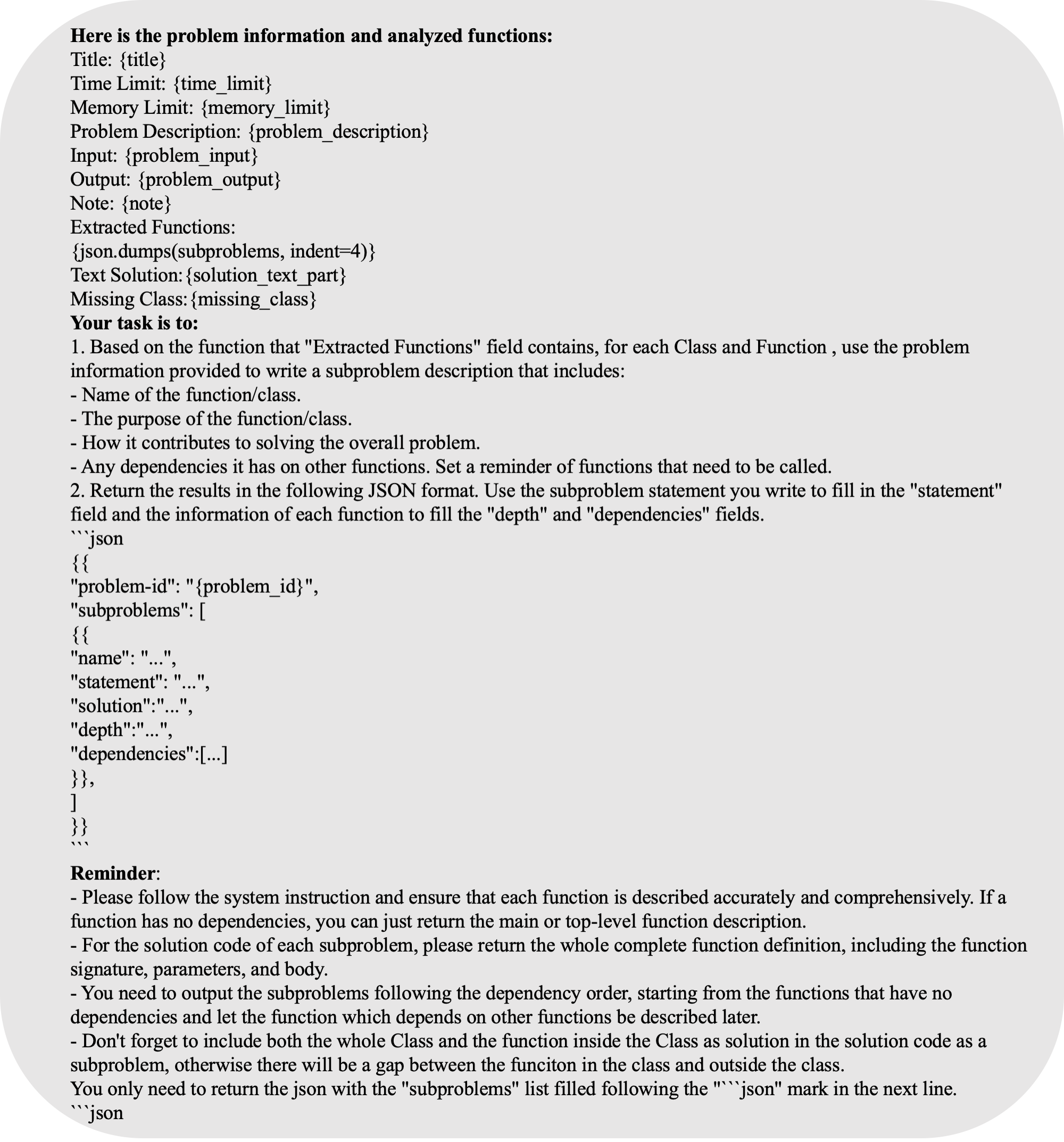}
    \caption{The prompt template used for generating natural language description for each subproblem in stage III.}
    \label{fig:Stage IV prompt template}
\end{figure*}

\begin{figure*}
    \centering
    \includegraphics[width=1\linewidth]{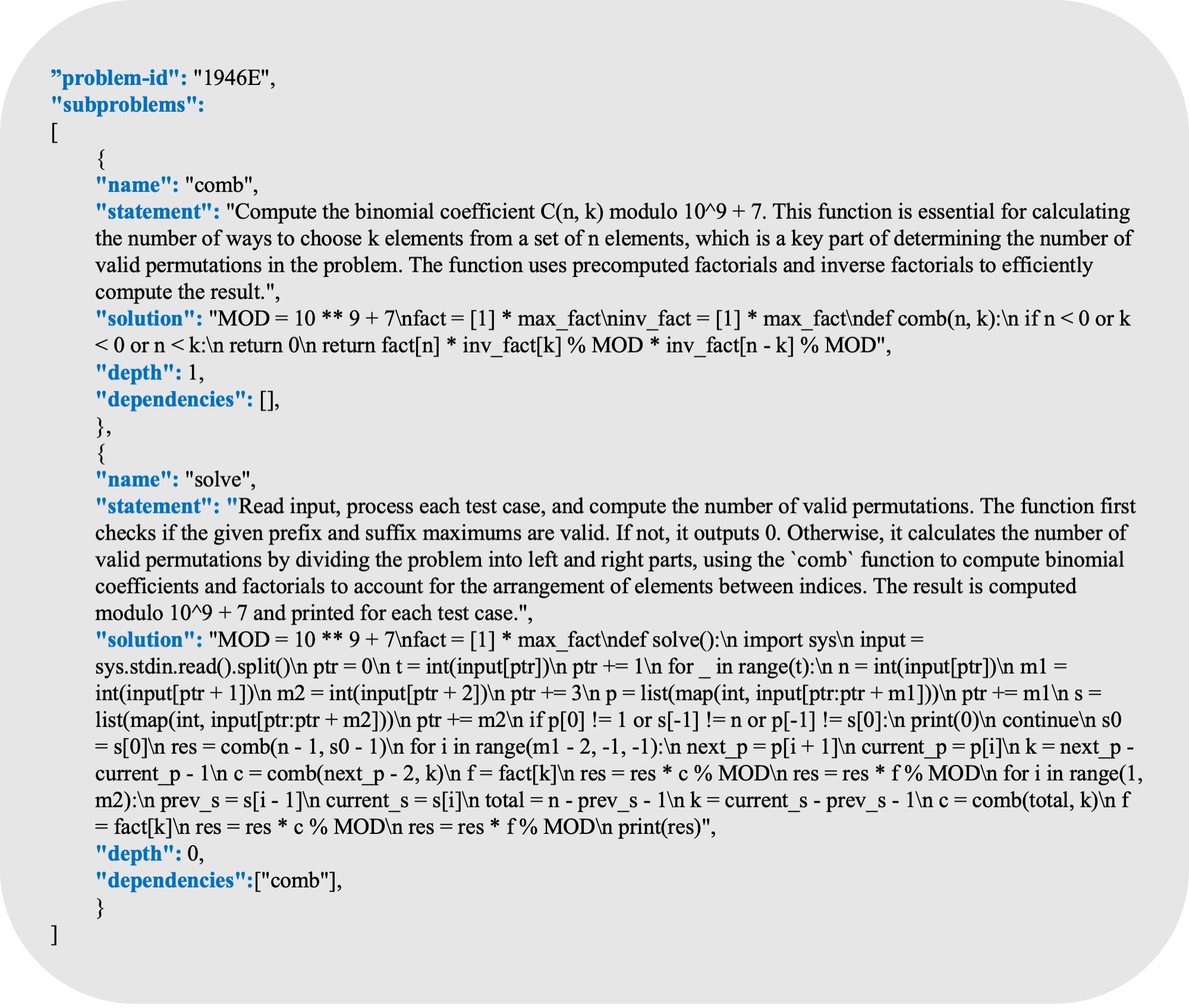}
    \caption{An example of subproblems we obtained in stage IV.The solution code of 1946E contains a \texttt{comb} function which serve as the basic tool and is reused in \texttt{solve} function to address the whole problem.It's obvious that in its AST, the \texttt{comb} function is the leave node in depth 1 and the \texttt{solve} function is the root node in depth 0.}
    \label{fig:Stage IV demo}
\end{figure*}

While the parsing process is automatically, a LLM is still needed for generating natural language description for each subproblem.The prompt template is shown in Figure \ref{fig:Stage IV prompt template}
A example of subproblems in shown in Figure \ref{fig:Stage IV demo}.

\subsection{Detail of Test Cases Generation}
\label{sec: Detail of Test Cases Generation}
Overall, CodeFlowBench’s test suite is composed of two parts:
\begin{itemize}
    \item \textbf{Top-level function tests}
For the final subproblem—the top-level function (e.g., \texttt{main} or \texttt{solve}) that handles overall input and output—we use the public test cases provided by the Codeforces platform.
    \item \textbf{Subproblem function tests}
For every other subproblem (i.e. functions invoked by higher-level code), we wrap each function call in a helper that redirects \texttt{stdin} and \texttt{stdout} to an internal buffer and records the resulting I/O. This can generate redundant calls for the same function, so we apply two safeguards to keep the test suite concise: deduplication of identical test cases and limiting the total number of test cases per function.These measures ensure comprehensive coverage without unnecessary duplication or excessive test-case volume.
\end{itemize}
Such method can ensure the test case generation pipeline to be automatic and the case itself to be correct.
\subsection{Ablation Study of LLM usage}
\label{sec: Ablation Study of LLM usage}
First, to check its influence on the content and style of Subproblem Descriptions, we compared Deepseek-V3 against GPT-4o-mini, Gemini-2.0-Flash and o3-mini. As shown in Table \ref{tab:desc_robustness}, the high semantic overlap and consistent stylistic metrics indicate that descriptions are primarily driven by the underlying code logic rather than model-specific habits. The main content of subproblems is not highly influenced by model choice. 

Second, to check its influence on the structure of LLM-generated solution code, we re-generated solutions using o3-mini, Claude-3.7-Sonnet, and Gemini-2.5-Pro. The resulting complexity metrics (Table \ref{tab:structure_metrics}) were identical across all models, confirming that the model produces almost identical decomposition results, indicating very minor bias between different LLMs.

Finally, to verify that our subproblems are natural and representative of human problem-solving, we conducted a Human-LLM cross-validation by constructing a creteria including five aspects. As shown in Table \ref{tab:quality_eval}, both human and model judges consistently rated the subproblems as high-quality, which  confirms that our automated pipeline produces professional, human-aligned problem decompositions.

Prompts of Subproblem Description Quality Validation is shown as follows:

\begin{tcolorbox}[
    colframe = gray,       
    colback = gray!5!white,             
    coltitle = white,                   
    coltext = black,                    
    fonttitle = \bfseries,              
    title = {Subproblem Description Quality Validation},  
    boxrule = 1pt,                      
    arc = 2mm,                          
    width = \linewidth,                 
    left = 7pt,                         
    right = 7pt,                        
    top = 5pt,                          
    bottom = 5pt                        
]
\fontsize{8.5pt}{10pt}\selectfont
You are an expert evaluator of programming task subproblems.\\
You will receive a subproblem (including the statement, solution code, and test cases). Evaluate whether the statement adequately meets each of the following five criteria. For each criterion, assign **1** if it is satisfied, or **0** if it is not. The total possible score is thus 5.\\

1. Clarity: Is the statement smoothly and understandably expressed, allowing a reader to grasp the task goal quickly?\\
2. Completeness: Does the description include all key elements needed to accomplish the subtask?\\
3. Accuracy: Is the description free of ambiguity or logical errors, and does it match the problem requirements and the provided code?\\
4. Feasibility: Can an engineer unambiguously determine and implement the required functionality—and pass functional tests—based solely on this description?\\
5. Professionalism: Does it use accurate, domain-appropriate terminology and a style fitting technical norms of coding tasks?\\

Return your result **only** as a JSON dictionary with these five keys and values of 0 or 1.\\

Please do not be overly strict. Assign a score of 1 to a criterion if it is even partially satisfied, allowing for minor imperfections. Only give a score of 0 if there are major or fundamental issues. Furthermore, ensure that the overall score for any subproblem is not lower than 2.\\
Here is the subproblem:{content}\\

Only return the JSON dictionary and remove the ```json note, nothing else.
\end{tcolorbox}

\begin{table}[h]
    \centering
    \caption{Robustness of Subproblem Descriptions (Ref: Deepseek-V3). \textbf{Left}: Semantic Consistency; \textbf{Right}: Stylistic Similarity.}
    \label{tab:desc_robustness}
    \small
    \resizebox{\columnwidth}{!}{
    \begin{tabular}{l|cc|ccc}
        \toprule
        \multirow{2}{*}{\textbf{Model}} & \multicolumn{2}{c|}{\textbf{Semantic}} & \multicolumn{3}{c}{\textbf{Stylistic}} \\
         & \textbf{ROUGE-1} & \textbf{BERT} & \textbf{Len. Ratio} & \textbf{Marker Diff.} & \textbf{Error Diff.} \\
        \midrule
        GPT-4o-mini & 0.582 & 0.725 & 0.975 & 0.026 & 0.367 \\
        Gemini-2.0-Flash & 0.600 & 0.734 & 1.141 & 0.019 & 0.728 \\
        o3-mini & 0.554 & 0.692 & 1.181 & 0.124 & 1.350 \\
        \bottomrule
    \end{tabular}
    }
\end{table}

\begin{table}[h]
    \centering
    \caption{Structural Complexity Metrics Across Models}
    \label{tab:structure_metrics}
    \small
    \begin{tabular}{lcc}
        \toprule
        \textbf{Model} & \textbf{Avg. Turns} & \textbf{Avg. Depth} \\
        \midrule
        o3-mini & 2.08 & 1.94 \\
        Claude-3.7-Sonnet & 2.08 & 1.94 \\
        Gemini-2.5-Pro & 2.08 & 1.94 \\
        \bottomrule
    \end{tabular}
\end{table}

\begin{table}[h]
    \centering
    \caption{Human Evaluation of Subproblem Quality. Metrics include Clarity (Cla.), Completeness (Com.), Accuracy (Acc.), Feasibility (Fea.), and Professionalism (Pro.).}
    \label{tab:quality_eval}
    \small
    \resizebox{\columnwidth}{!}{
    \begin{tabular}{lcccccc}
        \toprule
        \textbf{Source} & \textbf{Avg. Rating} & \textbf{Cla.} & \textbf{Com.} & \textbf{Acc.} & \textbf{Fea.} & \textbf{Pro.} \\
        \midrule
        o3-mini & \textbf{4.222} & 0.981 & 0.604 & 0.955 & 0.797 & 0.986 \\
        Claude-3.7 & 4.038 & 0.889 & 0.745 & 0.745 & 0.668 & 0.990 \\
        Human & 4.202 & 0.981 & 0.659 & 0.904 & 0.726 & 0.933 \\
        \bottomrule
    \end{tabular}
    }
\end{table}

\subsection{Compute Cost Estimation}
\label{sec: cost estimation}
The cost of using LLM in our data curation pipeline is estimated as below:
\begin{itemize}
    \item Generating Solution Code:
On average, each problem requires approximately 3K input tokens and 5K output tokens (including reasoning). The official Deepseek-R1 API is currently 4 CNY per 1M input tokens and 16 CNY per 1M output tokens. Therefore, the total cost for 5K+ problems is approximately 450 CNY.
    \item Generating Subproblem Description:
On average, each problem requires approximately 3K input tokens and 1K output tokens. The official Deepseek-V3 API is currently 2 CNY per 1M input tokens and 8 CNY per 1M output tokens. Therefore, the total cost for 5K+ problems is approximately 70 CNY.
\end{itemize}

The cost of processing the initial batch of 5K+ problems is manageable and also a one-time effort that does not need to be repeated. Future updates to the benchmark will involve only newly published problems and will therefore be much smaller in scale and less resource-intensive.

\section{CodeFlowBench-Repo Dataset}
\label{sec: codeflowbench-domain dataset}
In this section, we provide a detailed overview of the \textit{CodeFlowBench-Repo} dataset.

\subsection{DomainEval}
\label{sec: domaineval}
\paragraph{Overview of DomainEval}

DomainEval is an auto-constructed benchmark specifically designed to evaluate LLMs' coding capabilities across diverse real-world domains. Unlike general-purpose benchmarks that focus on common algorithmic tasks, DomainEval targets six distinct domains: \textit{Computation, Network, Basic, System, Visualization, and Cryptography}. The dataset is curated from representative open-source GitHub repositories that possess high community recognition (typically over 100 stars) to ensure code quality.

Structurally, each subject in DomainEval consists of three essential components: a natural language instruction, a reference solution (target function), and a suite of valid test cases extracted via a test-method matching strategy. Crucially, to ensure executability, DomainEval preserves the necessary context required to run the target code. This design ensures that the tasks are not isolated snippets but are deeply embedded in the dependency structures of authentic software projects.

\paragraph{Rationale for Source Selection.}
The selection of DomainEval as the foundation for \textit{CodeFlowBench-Domain} is a deliberate design choice driven by two key factors:

\textit{1. Authenticity across Specialized Domains.}
First, DomainEval provides a rigorous source of ecologically valid code samples. By extracting problems from mature, production-level repositories, it captures the authentic coding patterns that professional developers actually employ. Furthermore, its explicit coverage of specialized domains like \textit{Cryptography} and \textit{System} areas where general LLMs often struggle allows our benchmark to assess Dependency Awareness in contexts that require precise API knowledge, effectively complementing the algorithmic logic focus of standard competition datasets.

\textit{2. Alignment with Constructive Codeflow Paradigm.}

While many repository-level benchmarks focus on \textit{maintenance} tasks such as issue resolution or debugging within existing codebases, CodeFlowBench aims to evaluate the constructive aspect of software engineering, namely to build complex functionality from the ground up. 
DomainEval focuses on function generation within a rich context. This format is the ideal input for our pipeline, allowing us to decompose a monolithic, dependency-heavy function into a step-by-step, iterative construction process. By leveraging DomainEval, we effectively demonstrate that the iterative codeflow paradigm is adaptable to building functional modules in authentic software libraries, validating our pipeline's effectiveness beyond algorithmic puzzles.

\subsection{Dataset Construction}
\label{sec: domain dataset construction}
\begin{figure*}
    \centering
    \includegraphics[width=1\linewidth]{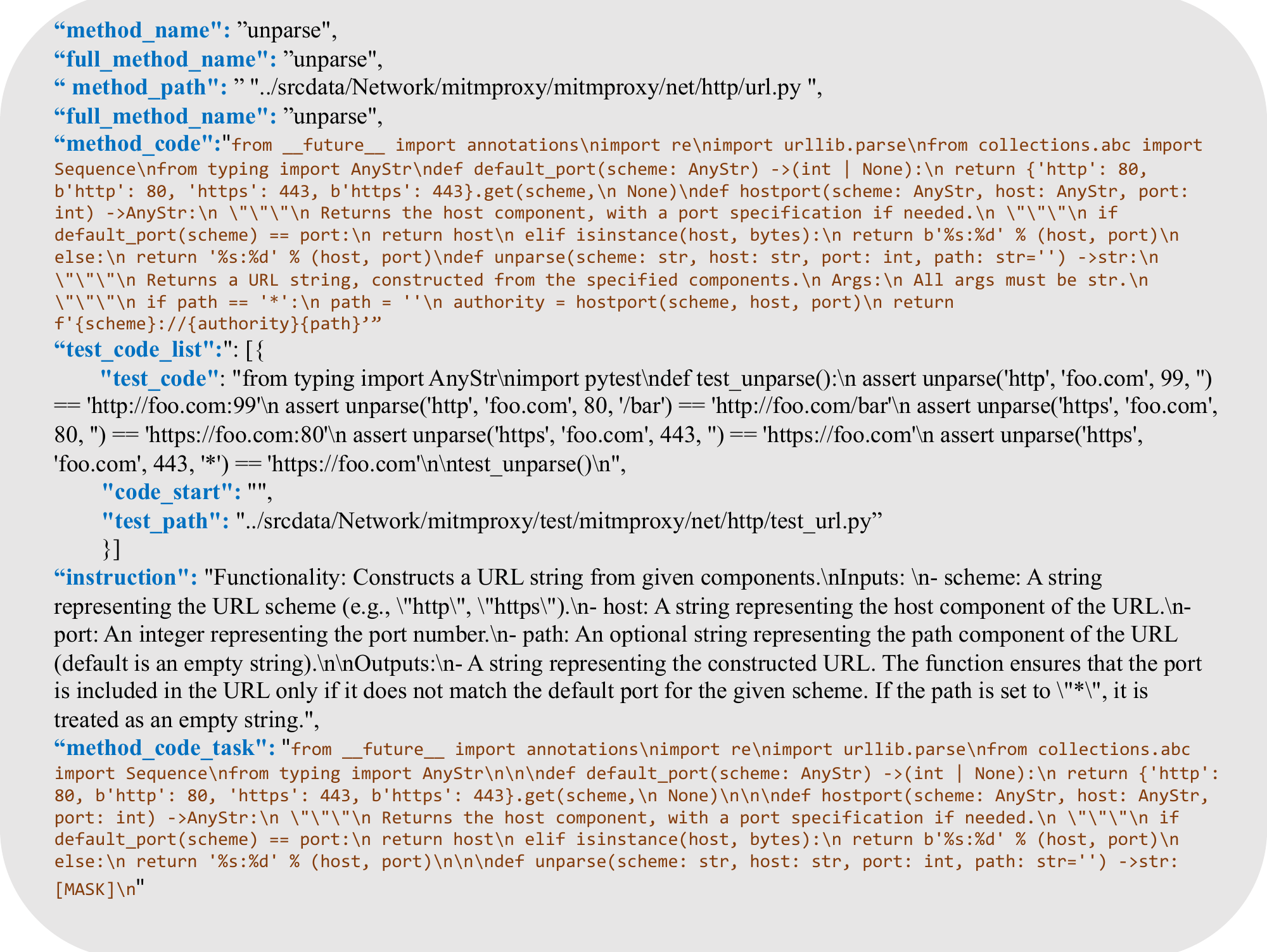}
    \caption{An example problem in DomainEval. We use the "method\_code" attribute to generate subproblem solution code and use test case in "test\_code" to generate subproblems' test cases.}
    \label{fig:DomainEval}
\end{figure*}

The construction of the CodeFlowBench-Repo subset adapts the Subproblem Generation Phase from the CodeFlowBench-Comp pipeline. A example of problems in DomainEval is shown in Fig. \ref{fig:DomainEval}

\begin{figure*}
    \centering
    \includegraphics[width=0.8\linewidth]{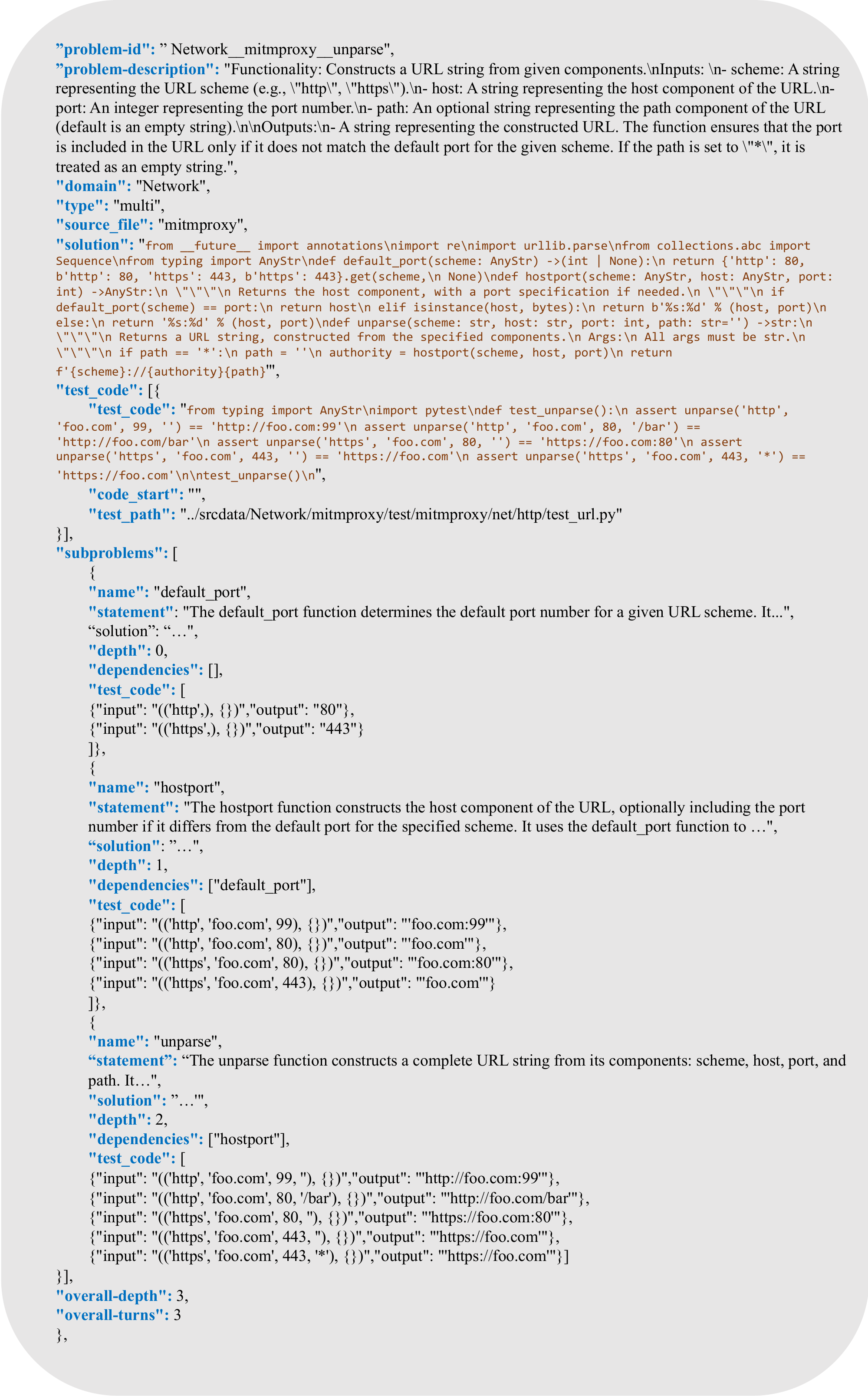}
    \caption{An example problem in CodeFlowBench-Repo, which has a shared form of problems in CodeFlowBench-Repo.}
    \label{fig:CodeFlowBench-Domain}
\end{figure*}

Since DomainEval already provides high-quality reference functions as ground truth, we omit the verification step and proceed directly to dependency analysis. We remove problems which only include single-turn function from the dataset and for the remaining, we parse the dependency relationship of the reference code to identify internal function calls and decompose them into iterative subproblems. Similarly, we then call LLM to generate problem description for subproblems. Finally, we use test cases of the top-level function to generate test cases of subproblems. An example is shown in Fig. \ref{fig:CodeFlowBench-Domain}. 

To ensure robust evaluation, we parse the import statements from the original test files and utilize Docker containers to encapsulate the complex package dependencies required by these real-world repositories. This ensures a compatible and reproducible execution environment for all generated subproblems.

\subsection{Dataset Statistics}
\label{sec: Domain Stats}
Since our pipeline explicitly looks for internal function reuse to create multi-turn steps, the API-centric nature of the Computation domain resulted in empty dependency trees. Consequently, our selection was restricted to the remaining 749 problems from the other five domains. After further filtering, only 70 high-quality multi-turn problems survived the pipeline. To ensure a balanced evaluation, we augmented this set by add 12 single-turn problems each domain and  curate the final CodeFlowBench-Repo.

\begin{figure*}
    \centering
    \includegraphics[width=1\linewidth]{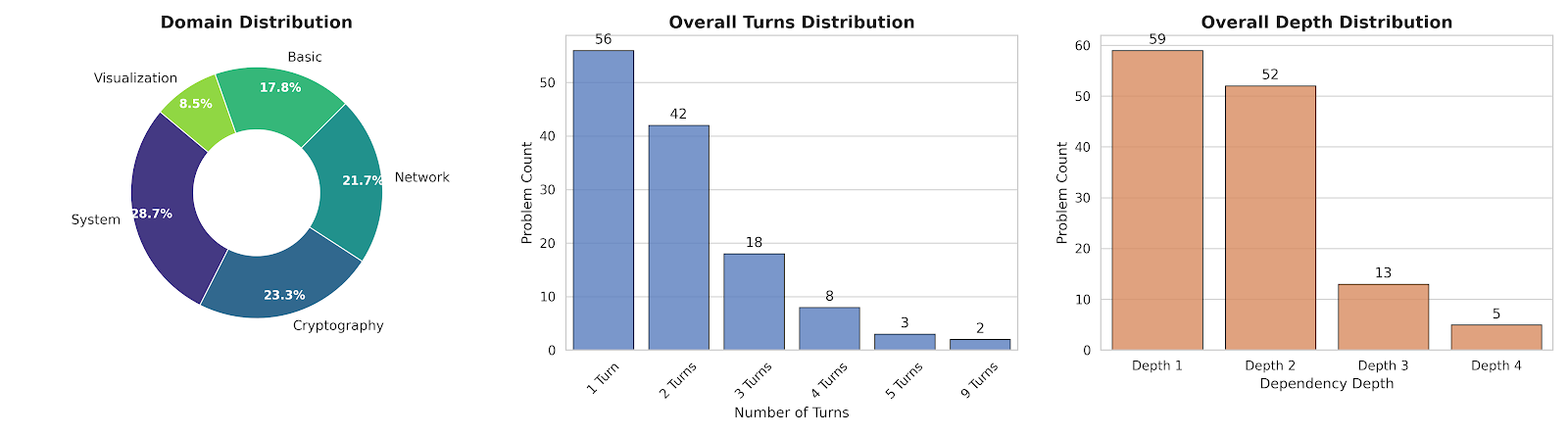}
    \caption{Statistics of CodeFlowBench-Repo. Left: The diversity of domains, with a focus on System and Cryptography
tasks. Right: Distributions of interaction turns and dependency depths,
highlighting the structural complexity.}
    \label{fig:domain-stat}
\end{figure*}

As illustrated in Fig. \ref{fig:domain-stat}, the dataset features a balanced distribution across five distinct domains: System (28.7\%) and Cryptography (23.3\%) constitute the majority, followed by Network (21.7\%), Basic utilities (17.8\%), and Visualization (8.5\%). This distribution ensures the benchmark covers a wide spectrum of software engineering scenarios, from low-level OS interactions to high-level data plotting.

Structurally, the dataset presents significant complexity beyond atomic code generation. The distribution of overall turns and overall depths is shown in Fig. \ref{fig:domain-stat}.  The distribution of overall turns reveals that 56.6\% of the problems require multi-turn iterative generation, forcing models to maintain reasoning continuity over long horizons. Furthermore, the dependency depth distribution confirms the intricate nature of these tasks: while 45.7\% of problems act as foundational nodes (Depth 1), the remaining 54.3\% involve deeper dependency chains (Depth 2--4), requiring the model to correctly implement and call prerequisite functions within a hierarchical context. 


\section{Mathematical Expression of $PD@k$\& $APD@k$}
\label{sec: mathematical expression}
For a given problem, we define $PD@k$ as the expected maximum pass depth over $k$ independent trials of the model.  Directly using only those $k$ results leads to high variance, just as with the $pass@k$ metric.  By analogy to the unbiased estimator for $pass@k$, which leverages $n$ trials ($n>k$) to reduce variance, we derive a similar estimator for $PD@k$.

Let the pass depths from $n$ trials be:
\[
\{d_1, d_2, \dots, d_n\},
\]
and let
\[
d_{(1)} \le d_{(2)} \le \cdots \le d_{(n)}
\]
denote these depths in ascending order.  Then an unbiased estimator for $PD@k$ is
\begin{equation}
PD@k \;=\; 
\sum_{j=k}^{n} d_{(j)} \,\frac{\binom{j-1}{k-1}}{\binom{n}{k}}.
\end{equation}

To see how this arises, consider sampling a random subset of size $k$ from the $n$ depths.  Let $M \;=\; \max\{\,d_{i_1}, d_{i_2}, \dots, d_{i_k}\}$ be the maximum depth in that subset. An unbiased estimator of $M$ is $\mathbb{E}[M]$. By construct
$\mathbb{E}[M] \;=\; PD@k$
,this estimator is \emph{unbiased} since its expected value exactly equals the true expected maximum depth. Consider all possible value of $M$, we have:
\begin{equation}
\mathbb{E}[M]
= \sum_{j=k}^{n} d_{(j)}\,P\bigl(M = d_{(j)}\bigr).
\end{equation}
The probability that $M=d_{(j)}$ equals the probability of choosing $d_{(j)}$ along with $k-1$ depths from the first $j-1$ smaller values:
\begin{equation}
P\bigl(M = d_{(j)}\bigr)
= \frac{\displaystyle\binom{j-1}{k-1}}{\displaystyle\binom{n}{k}}.
\end{equation}
Combining (3) and (4) immediately recovers (1).

Finally, we define
\begin{equation}
APD@k = \frac{1}{|\mathcal{P}|}\sum_{p\in\mathcal{P}} PD@k(p)  
\end{equation}
the average $PD@k$ over the set $\mathcal{P}$ of all problems.

\section{Experiment}
\subsection{Experiment Setting}
\label{sec: Experiment Setting}
Due to the substantial size of the total question pool ($N=5258$), we implemented stratified sampling with proportional allocation across overall-depth categories to select 1,000 test samples, as detailed in Table~\ref{tab:sampling1}. To validate sampling quality, we conducted a $\chi^2$ test comparing the overall-turn distributions between the population and sampled data (Table~\ref{tab:sampling2}). The statistical analysis yielded a p-value of 0.1246, indicating no significant difference ($\alpha = 0.05$) in distribution patterns. This confirms the representativeness of our sampling strategy and ensures the validity of subsequent analytical outcomes.
\begin{table*}[htbp]
\centering
\caption{Comparison of Overall-Depth Proportions Between the Population and the Sample}
\label{tab:sampling1}
\begin{tabular}{*{5}{c}}
\toprule
\textbf{Overall-Depth} & 
\textbf{\# Population} & 

\textbf{Pop. Proportion (\%)} & 
\textbf{\# Sample} & 
\textbf{Sam. Proportion (\%)} \\
\midrule
1 & 1,488 & 28.3 & 283 & 28.3 \\
2 & 2,751 & 52.3 & 523 & 52.3 \\
3 & 870 & 16.5 & 165 & 16.5 \\
4 & 125 & 2.4 & 24 & 2.4 \\
$\geq$5 & 24 & 0.5 & 5 & 0.5 \\
\midrule
\textbf{Total} & \textbf{5,258} & \textbf{100.0} & \textbf{1,000} & \textbf{100.0} \\
\bottomrule
\end{tabular}
\end{table*}

\begin{table*}[htbp]
\centering
\caption{Comparison of Overall-Turn Proportions Between the Population and the Sample}
\label{tab:sampling2}
\begin{tabular}{*{5}{c}}
\toprule
\textbf{Overall-Turn} 
& \textbf{\# Population} & 
\textbf{Pop. Proportion (\%)} & 
\textbf{\# Sample} & 
\textbf{Sam. Proportion (\%)} \\
\midrule 
1 & 1,488 & 28.3 & 283 & 28.3 \\
2 & 2,158 & 41.0 & 437 & 43.7 \\
3 & 990 & 18.8 & 188 & 18.8 \\
4 & 402 & 7.6 & 58 & 5.8 \\
5 & 149 & 2.8 & 21 & 2.1 \\
6 & 46 & 0.9 & 6 & 0.6 \\
$\geq$7 & 25 & 0.5 & 7 & 0.7 \\
\midrule
\textbf{Total} & \textbf{5,258} & \textbf{100.0} & \textbf{1,000} & \textbf{100.0} \\
\bottomrule
\end{tabular}
\end{table*}

To ensure the reproducibility of our results, we detail the inference hyperparameters in Table~\ref{tab:hyperparameters}. 

\begin{table*}[ht]
\centering
\caption{Inference Hyperparameters for Evaluation}
\label{tab:hyperparameters}
\begin{tabular}{@{}lccc@{}}
\toprule
\textbf{Hyperparameter} & \textbf{API} & \textbf{Local} & \textbf{Description} \\ \midrule
Temperature & 0.6 & 0.6 & Balances creative reasoning with logical consistency. \\
Top-p & 1.0 & 1.0 & No nucleus sampling truncation applied. \\
Max Tokens & 5,000 & 10,000 & Maximum tokens generated per subproblem turn. \\
Frequency Penalty & 0.0 & 0.0 & Prevents repetitive token generation. \\ \bottomrule
\end{tabular}
\end{table*}

\subsection{Instruction Templates}
\label{sec:Instruction Templates}

In the instruction template for multi-turn testing, we implemented a set of basic heuristics to adapt to different problem types. For example, if a question has no dependencies, we omit the “\#\# Dependency information” section. If no prior code is provided—which is common when the overall depth is 1—we exclude the “\#\# Pre-implemented functions” section. Furthermore, if it is the final turn of the code, we append the following snippet:

\begin{lstlisting}[language=python, basicstyle=\ttfamily\footnotesize]
import sys
def {name}():
    input = sys.stdin.read().split()
\end{lstlisting}
Similarly, if the problem does not have any dependencies, we will also omit the section related to \{dependencies\}.

\begin{tcolorbox}[
    colframe = gray,       
    colback = gray!5!white,             
    coltitle = white,                   
    coltext = black,                    
    fonttitle = \bfseries,              
    title = {Multi-turn Test},  
    boxrule = 1pt,                      
    arc = 2mm,                          
    width = \linewidth,                 
    left = 7pt,                         
    right = 7pt,                        
    top = 5pt,                          
    bottom = 5pt                        
]
\fontsize{8.5pt}{10pt}\selectfont
You are a Programming Expert. You always provide correct and reliable code solutions. You will be provided with the background of the whole problem, a programming problem and may also some pre-implemented functions.If pre-implemented functions provided, you need to call the pre-implemented functions and write a new function to solve the problem.\\
\#\# Background of the whole problem:\\
\{problem\_description\}\\
\#\# Problem Description:
You need to complete {name} function.\\
\{statement\}\\
\#\# Dependency information:\\
To solve the problem, you need to utilize the \#\# Pre-implemented functions \{dependencies\} provided.\\
\#\# Pre-implemented functions:\\
\{history\}\\
\#\# Guidelines:\\
- Ensure the function is executable and meets the requirement.\\
- Handle \#\# Dependency information correctly.\\
- Provide clear and concise comments to explain key parts of the code. \\
Return your response by filling the function body following the function signature provided. Just generate the function and don't output any examples.\\
\end{tcolorbox}

\label{app:instructions2}
\begin{tcolorbox}[
    colframe = gray,       
    colback = gray!5!white,             
    coltitle = white,                   
    coltext = black,                    
    fonttitle = \bfseries,              
    title = {Single-turn Test},  
    boxrule = 1pt,                      
    arc = 2mm,                          
    width = \linewidth,                 
    left = 7pt,                         
    right = 7pt,                        
    top = 5pt,                          
    bottom = 5pt                        
]
\fontsize{8.5pt}{10pt}\selectfont
You are a Programming Expert. You always provide correct and reliable code solutions. You are required to solve a problem which consists of multiple subproblems, each with its own requirements. 
You will be provided with the background of the problem and description of all subproblems. You need to generate the complete implementations for all subproblems in a single response.  
The response for the final subproblem will be tested using stdin and stdout. Ensure the corresponding code meet this requirement.\\
\#\# Background of the whole problem:\\
\{problem\_description\}\\
\#\# Problem Description:\\
\{combined\_subproblem\_description\}\\
\#\# Subproblem \{name\} \\
\# Description:\\
You need to complete \{name\} function.\\
\{statement\} \\
To solve the problem, you need to utilize your pre-implemented function \{dependencies\}.\\
\#\# Guidelines:
- Ensure that all functions are executable and meet their respective requirements.\\
- For each subproblem, correctly handle any dependency information.\\
- Provide clear and concise comments explaining the key parts of the code.\\
- For the last subproblem {name}, please use 'import sys\textbackslash ndef \{name\}():\textbackslash n~~~~input = sys.stdin.read().split()\textbackslash n" as the beginning.\\
Return your response by generating all functions in a single code block. \\
\end{tcolorbox}


\subsection{Statistic Rigor}
\label{sec:Statistic Rigor}

To ensure the statistical rigor of our empirical findings and verify that the observed performance gaps are not merely artifacts of data sampling variance, we re-evaluated our main results by constructing 95\% confidence intervals.

Given the nature of pass rates (Pass@k) and Average Pass Depth (APD@k), which may not strictly follow a normal distribution, we employed non-parametric bootstrapping. For each dataset and model combination, we generated $N=1000$ bootstrap resamples (with replacement) from the original test sets. We calculated the metrics for each resample to construct a bootstrap distribution. The standard error (SE) was derived from the standard deviation of this distribution, which was then used to approximate the 95\% confidence interval.

Table \ref{tab:bootstrap_ci} presents the performance of representative models across different tiers, augmented with the computed bootstrap standard errors.

\begin{table*}[htbp]
\centering
\caption{Main Results with Bootstrap Confidence Intervals ($\pm$ Standard Error, 95\% CI via 1000 resamples). Pass@1 values are reported as percentages.}
\label{tab:bootstrap_ci}
\small 
\setlength{\tabcolsep}{4pt} 
\begin{tabular}{lcccccc}
\toprule
\multirow{2}{*}{\textbf{Model}} & \textbf{Multi-Turn} & \textbf{Single-Turn} & \multicolumn{4}{c}{\textbf{Overall APD@1 ($\pm$ SE) across Depths}} \\ \cmidrule(l){4-7}
 & \textbf{(Pass@1 \%)} & \textbf{(Pass@1 \%)} & \textbf{Depth 1} & \textbf{Depth 2} & \textbf{Depth 3} & \textbf{Depth 4} \\ \midrule
\multicolumn{7}{c}{\textit{\textbf{CodeFlowBench-Comp}}} \\ \midrule
GPT-5 & 26.5 $\pm$ 1.4 & 37.6 $\pm$ 1.7 & 0.185 $\pm$ 0.023 & 0.866 $\pm$ 0.039 & 1.306 $\pm$ 0.092 & 1.542 $\pm$ 0.284 \\
Gemini-3-flash-thinking & 48.4 $\pm$ 1.6 & 65.5 $\pm$ 1.5 & 0.445 $\pm$ 0.030 & 1.303 $\pm$ 0.040 & 1.625 $\pm$ 0.102 & 1.625 $\pm$ 0.321 \\
Claude-4.5-Sonnet & 31.6 $\pm$ 1.5 & 47.9 $\pm$ 1.6 & 0.222 $\pm$ 0.024 & 1.008 $\pm$ 0.040 & 1.377 $\pm$ 0.088 & 1.521 $\pm$ 0.248 \\
Qwen3-Coder-30B-A3B & 11.6 $\pm$ 1.0 & 26.0 $\pm$ 1.3 & 0.100 $\pm$ 0.018 & 0.458 $\pm$ 0.032 & 0.681 $\pm$ 0.070 & 0.917 $\pm$ 0.194 \\ \midrule
\multicolumn{7}{c}{\textit{\textbf{CodeFlowBench-Repo}}} \\ \midrule
GPT-5 & 34.7 $\pm$ 4.2 & 48.8 $\pm$ 4.3 & 0.483 $\pm$ 0.066 & 1.020 $\pm$ 0.110 & 1.385 $\pm$ 0.340 & 3.200 $\pm$ 0.429 \\
Gemini-3-flash-thinking & 33.9 $\pm$ 4.2 & 56.7 $\pm$ 4.4 & 0.500 $\pm$ 0.066 & 1.020 $\pm$ 0.113 & 1.308 $\pm$ 0.338 & 3.400 $\pm$ 0.358 \\
Claude-4.5-Sonnet & 33.1 $\pm$ 4.2 & 56.7 $\pm$ 4.3 & 0.448 $\pm$ 0.064 & 1.039 $\pm$ 0.111 & 1.615 $\pm$ 0.323 & 3.200 $\pm$ 0.436 \\
Qwen3-Coder-30B-A3B & 23.6 $\pm$ 3.8 & 42.5 $\pm$ 4.2 & 0.414 $\pm$ 0.064 & 0.471 $\pm$ 0.100 & 1.385 $\pm$ 0.376 & 0.600 $\pm$ 0.282 \\ \bottomrule
\end{tabular}
\end{table*}
\section{Ablation Study}
\label{sec: Ablation Study}
\subsection{Empirical Validation of Software Quality}

While Pass@k metrics effectively capture the functional correctness of generated code, they fall short in assessing its maintainability and alignment with software engineering best practices. To comprehensively evaluate whether the codeflow paradigm intrinsically yields higher code quality than monolithic generation, we conducted an empirical ablation study focusing on structural and maintainability metrics.We utilized standard static analysis tools (e.g., \texttt{radon} for Python) to extract the following metrics from the generated solutions:
\begin{itemize}
    \item \textbf{Average Function Count (FC):} A proxy for modularity. A higher count in solving the same problem indicates a stronger inclination toward problem decomposition.
    \item \textbf{Average Function Length (SLOC):} Source Lines of Code per function. Shorter functions generally indicate better encapsulation and strict adherence to the Single Responsibility Principle (SRP).
    \item \textbf{Comment Density (CD):} The ratio of comment lines to total lines, reflecting the code's self-documentation capabilities.
    \item \textbf{Cyclomatic Complexity (CC):} A quantitative measure of the number of linearly independent paths through a program's source code. Lower CC values imply that the code is easier to test, debug, and maintain.
\end{itemize}

We sampled successful completions for problems requiring $T > 1$ turns across three leading models  under both multi-turn and single-turn settings. Table \ref{tab:code_quality} demonstrates the result. Models operating in the codeflow paradigm generated more functions per solution with significantly higher comment density, effectively breaking down complex logic into manageable units. Particularly for Gemini-3 and Claude-4.5, the multi-turn setting led to a notable reduction in both SLOC and Cyclomatic Complexity per function. This confirms that guiding models through iterative subproblems actively prevents the generation of convoluted, monolithic code.

\begin{table*}[htbp]
\centering
\caption{Maintainability and Complexity Metrics across Models (Multi-turn vs. Single-turn)}
\label{tab:code_quality}
\begin{tabular}{llcccc}
\toprule
\textbf{Model} & \textbf{Setting} & \textbf{Avg. FC ($\uparrow$)} & \textbf{Avg. SLOC ($\downarrow$)} & \textbf{CD ($\uparrow$)} & \textbf{CC ($\downarrow$)} \\ 
\midrule
\multirow{2}{*}{Gemini-3}   & Multi-Turn  & 3.53 & 15.14 & 37.53\% & 5.11 \\
                            & Single-Turn & 2.75 & 16.12 & 33.50\% & 5.21 \\ 
\midrule
\multirow{2}{*}{Claude-4.5} & Multi-Turn  & 2.98 & 14.41 & 33.04\% & 4.54 \\
                            & Single-Turn & 2.28 & 20.06 & 23.56\% & 6.21 \\ 
\midrule
\multirow{2}{*}{GPT-5}      & Multi-Turn  & 2.93 & 19.35 & 31.43\% & 6.62 \\
                            & Single-Turn & 2.21 & 16.17 & 23.22\% & 5.29 \\ 
\bottomrule
\end{tabular}
\end{table*}

Beyond static metrics, the fundamental advantage of the multi-turn CodeFlow paradigm lies in its workflow alignment. Monolithic single-turn generation creates a ``black-box'' scenario where errors in logic or integration are notoriously difficult to trace. In contrast, our paradigm enables developers to unit-test atomic modules at each turn. By isolating functionality, errors are constrained to the current working module, significantly reducing the cognitive load required for debugging and mirroring real-world Agile development practices.

\subsection{Analysis of Data Contamination Risks}
\label{sec:appendix_contamination}

Given that CodeFlowBench sources data from public platforms like Codeforces and GitHub, a natural concern arises regarding potential data contamination. To ensure that our benchmark evaluates true reasoning capabilities rather than the models' memorization of their training corpora, we address this issue through both structural methodology and empirical validation.

\paragraph{Structural Barrier Against Retrieval}
We emphasize that the CodeFlow task extends far beyond simple code retrieval. Our AST-based pipeline transforms monolithic solutions into constrained, dependency-aware sequences. To succeed, models must demonstrate deep contextual reasoning to adapt the overarching logic into specific subproblem signatures and strict modular boundaries. This structural transformation acts as a natural barrier; simply recalling the original training data or a monolithic script is insufficient to pass the multi-turn tests.

\paragraph{Chronological Contamination Check}
To provide empirical evidence, we conducted a chronological contamination check on CodeFlowBench-Comp. We evaluated model performance across three distinct chronological slices sampled from the most recent 1,000 problems. To ensure a fair and unbiased comparison, these slices were strictly matched in terms of average difficulty ratings, interaction turns, and dependency depth. Table \ref{tab:contamination} presents the Pass@1 and APD@1 scores for Gemini-3 and Claude-4.5 across these time periods. As shown, there is no significant or consistent performance degradation on newer problems (Slice 3) compared to older ones (Slice 1). In fact, Claude-4.5 even shows a slight performance increase on newer problems. This lack of negative chronological correlation strongly suggests that the models' performance is driven by their intrinsic reasoning and logical capabilities, rather than the memorization of specific problem instances.

\begin{table*}[htbp]
\centering
\caption{Model Performance Across Chronological Slices (Controlled for Difficulty, Depth, and Turn)}
\label{tab:contamination}
\begin{tabular}{llccc}
\toprule
\textbf{Model} & \textbf{Chronological Slice} & \textbf{ID Range} & \textbf{Pass@1} & \textbf{APD@1} \\ \midrule
\multirow{3}{*}{Gemini-3}   & Slice 1 (Oldest) & 1609 - 1826 & 0.68 & 1.64 \\
                            & Slice 2 (Medium) & 1829 - 1942 & 0.62 & 1.55 \\
                            & Slice 3 (Newest) & 1943 - 2057 & 0.64 & 1.59 \\ \midrule
\multirow{3}{*}{Claude-4.5} & Slice 1 (Oldest) & 1609 - 1826 & 0.47 & 1.19 \\
                            & Slice 2 (Medium) & 1829 - 1942 & 0.49 & 1.26 \\
                            & Slice 3 (Newest) & 1943 - 2057 & 0.52 & 1.35 \\ \bottomrule
\end{tabular}%

\end{table*}

\subsection{Analysis of Evaluation Protocol: Fail-Stop vs. Non-Stop}
\label{sec:appendix_evaluation_protocol}

A potential concern regarding our primary evaluation framework is whether the strict ``fail-stop'' mechanism overly penalizes models for trivial early-stage errors, thereby obscuring their capacity to solve complex logic in subsequent turns. To rigorously isolate the models' logical capabilities at each specific step without the compounding effect of earlier failures, we conducted an ablation study using a ``non-stop'' (Teacher-Forcing) evaluation variant.

\paragraph{Experimental Setup}
We sampled 100 problems and evaluated three leading models. In the non-stop mode, if a model fails a specific turn, we inject the ground-truth implementation for that subproblem into the context. This intervention provides the model with a ``clean slate'' to tackle the subsequent dependencies. To quantify this, we introduce the \textit{Absolute Solved Turns (AST)} metric. Unlike standard pass metrics, AST calculates the absolute number of subproblems a model solves correctly per problem, given that any failed preceding turns are seamlessly replaced with ground-truth implementations.

\paragraph{Results}
Table \ref{tab:non_stop} contrasts the AST scores under the standard fail-stop setting versus the non-stop setting. While the non-stop mode predictably yields slightly higher scores, the overall performance gains ($\Delta$) are remarkably marginal . This minimal variance provides a crucial insight: a model that fails an early, low-complexity helper function genuinely struggles with the underlying logic and typically lacks the reasoning depth required for the heavier dependencies in subsequent turns. It proves that cascading failures in CodeFlowBench are largely indicative of true capability boundaries rather than trivial syntax errors.

\begin{table*}[htbp]
\centering
\caption{Comparison of Absolute Solved Turns (AST) under Standard vs. Non-stop Evaluation}
\label{tab:non_stop}
{%
\begin{tabular}{lcccccc}
\toprule
\multirow{2}{*}{\textbf{Model}} & \multirow{2}{*}{\textbf{Fail-Stop}} & \multirow{2}{*}{\textbf{Non-Stop}} & \multicolumn{4}{c}{\textbf{Performance Gain ($\Delta$)}} \\ \cmidrule(l){4-7} 
 & & & \textbf{Overall} & \textbf{2-Turns} & \textbf{3-Turns} & \textbf{4-Turns} \\ \midrule
Gemini-3   & 1.19 & 1.35 & +0.16 & +0.08 & +0.35 & +0.25 \\
GPT-5      & 0.80 & 0.87 & +0.07 & +0.07 & +0.10 & +0.00 \\
Claude-4.5 & 0.79 & 0.87 & +0.08 & +0.09 & +0.22 & +0.11 \\ \bottomrule
\end{tabular}%
}
\end{table*}
\subsection{Sensitivity to Topological Ordering}
\label{sec:appendix_topological}

A potential concern regarding the evaluation stability is that different linearizations of the same non-linear dependency graph could lead to arbitrary fluctuations in the APD and APT metrics. 

\paragraph{Experimental Setup}
To empirically verify the stability of our metrics against graph linearization, we sampled 100 problems exhibiting non-linear dependency structures (i.e., those allowing multiple valid topological sorts). For each problem, we generated four distinct, valid topological linearizations and evaluated the models across these variants. 

\paragraph{Results}
The results, summarized in Table \ref{tab:topological}, demonstrate that our structural metrics are remarkably robust to task reordering. Statistical analysis reveals that the variance of Average Pass Turn (APT) across topological variants remains exceptionally low (with Standard Deviations $\approx 0.04$), while the fluctuation in Average Pass Depth (APD) is even more negligible. 

This quantitative stability confirms that both metrics serve as consistent and reliable indicators of model performance. Although different valid linearizations alter the immediate preceding context, they primarily shift the position of parallel subproblems without destabilizing the overall evaluation ranking or obscuring the models' true iterative capabilities.

\begin{table*}[htbp]
\centering
\caption{Comparison of APD and APT Metrics across Topological Variants}
\label{tab:topological}
\begin{tabular}{lcccccc}
\toprule
\multirow{2}{*}{\textbf{Variant}} & \multicolumn{2}{c}{\textbf{Gemini-3}} & \multicolumn{2}{c}{\textbf{GPT-5}} & \multicolumn{2}{c}{\textbf{Claude-4.5}} \\ \cmidrule(l){2-3} \cmidrule(l){4-5} \cmidrule(l){6-7}
 & \textbf{APD} & \textbf{APT} & \textbf{APD} & \textbf{APT} & \textbf{APD} & \textbf{APT} \\ \midrule
Variant 1 & 0.86 & 1.30 & 0.57 & 1.11 & 0.59 & 1.13 \\
Variant 2 & 0.86 & 1.33 & 0.59 & 1.22 & 0.63 & 1.18 \\
Variant 3 & 0.87 & 1.40 & 0.58 & 1.17 & 0.59 & 1.13 \\
Variant 4 & 0.84 & 1.40 & 0.56 & 1.10 & 0.56 & 1.07 \\ \midrule
\textbf{Mean (APT)} & \multicolumn{2}{c}{1.3575} & \multicolumn{2}{c}{1.1500} & \multicolumn{2}{c}{1.1275} \\
\textbf{Std. Dev. (APT)} & \multicolumn{2}{c}{0.0437} & \multicolumn{2}{c}{0.0461} & \multicolumn{2}{c}{0.0450} \\ \bottomrule
\end{tabular}
\end{table*}

\subsection{Error Case Study}
\label{sec: Error Case Study}
We have defined three typical error types for test in multi-turn pattern, in this part we will introduce several example to illustrate them.We have simplify and arrange model's outputs to make it clear to read. The content of "generated" field is model's output. The content of "harness\_result" field is the verification result by running it with test cases."1" denotes accepted, "0" denotes wrong answer and "wrong" denotes running error.

\begin{figure*}
    \centering
    \includegraphics[width=1\linewidth]{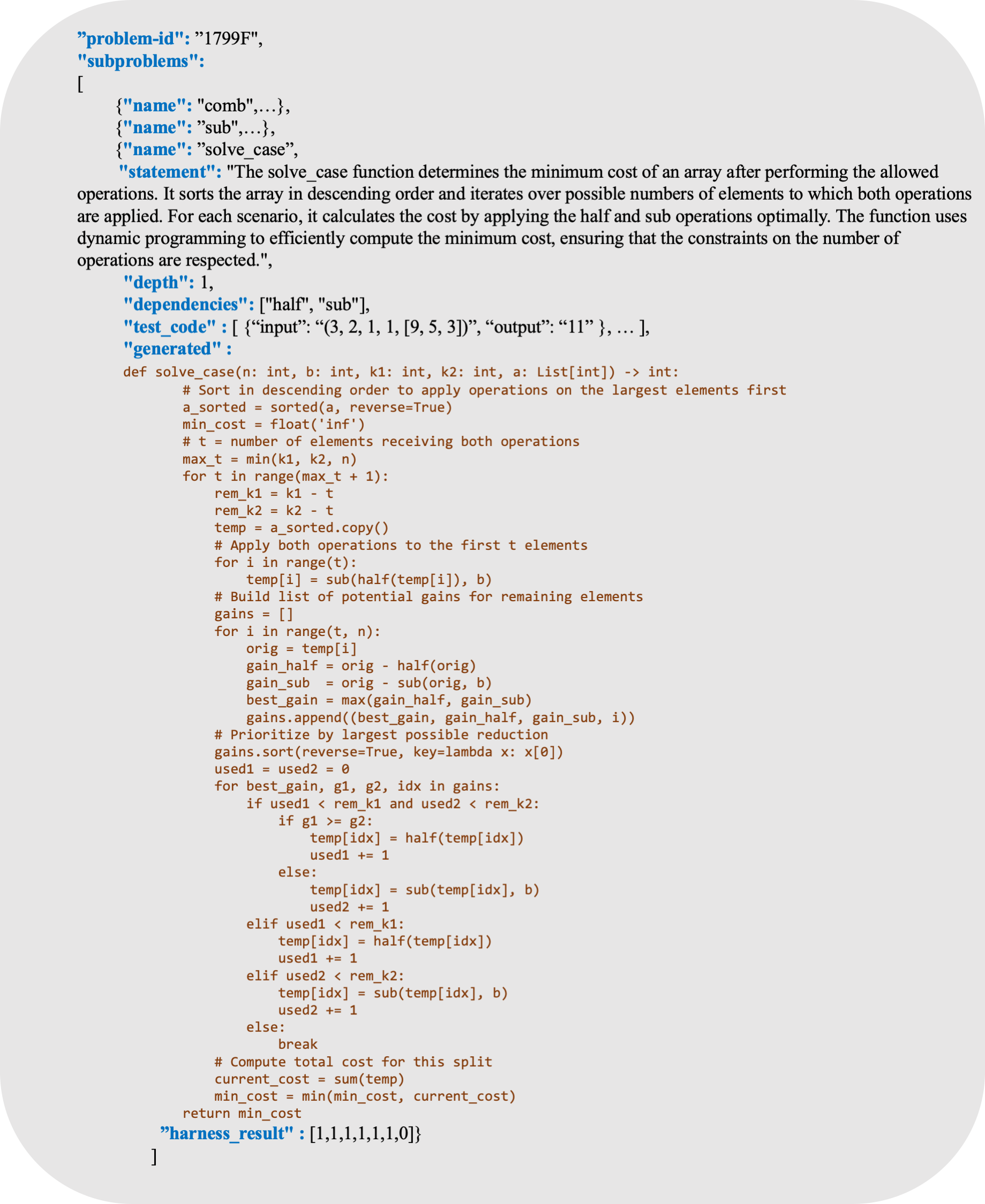}
    \caption{Example of an Incomplete Reasoning (IR) Error by Deepseek-V3.The original problem is \href{https://codeforces.com/problemset/problem/1799/F}{1799F}.The algorithm produced by the model fails to find the global optimum because it greedily sorts each element according to the single-step gain—choosing either the halving operation or the subtraction operation in isolation. In reality, achieving the global optimum requires a two-dimensional dynamic-programming solution that considers both operations jointly.}
    \label{fig:IR}
\end{figure*}
\begin{figure*}
    \centering
    \includegraphics[width=0.85\linewidth]{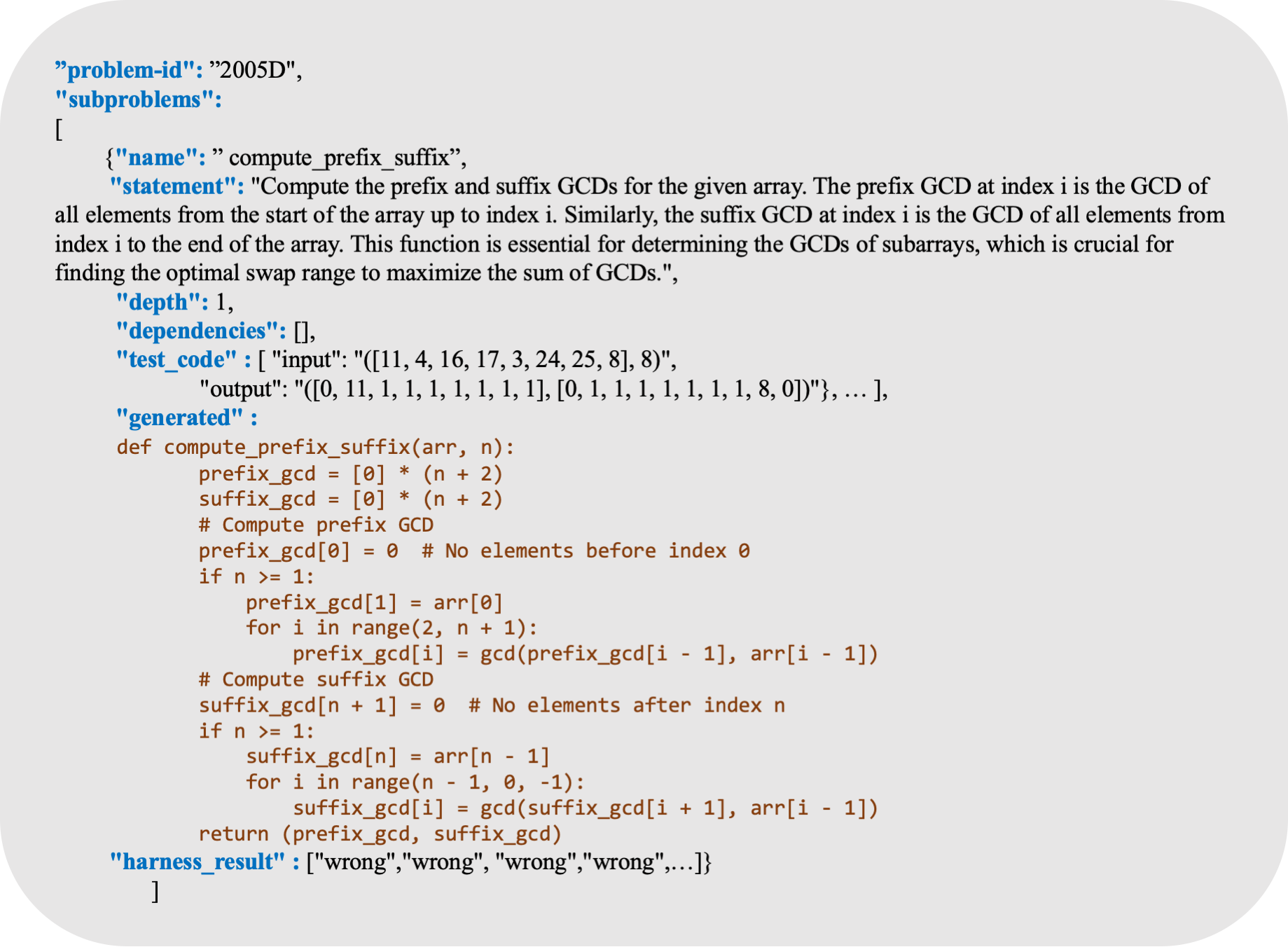}
    \caption{Example 1 of an Insufficient Globalization(IG) error by Deepseek-V3.The original problem is \href{https://codeforces.com/problemset/problem/2005/D}{2005D}.The model’s generated code omitted the import \texttt{math} statement, resulting in an error when calling \texttt{gcd}. This issue stems from improper handling of external imports.} 
    \label{fig:IG1}
\end{figure*}
\begin{figure*}
    \centering
    \includegraphics[width=0.85\linewidth]{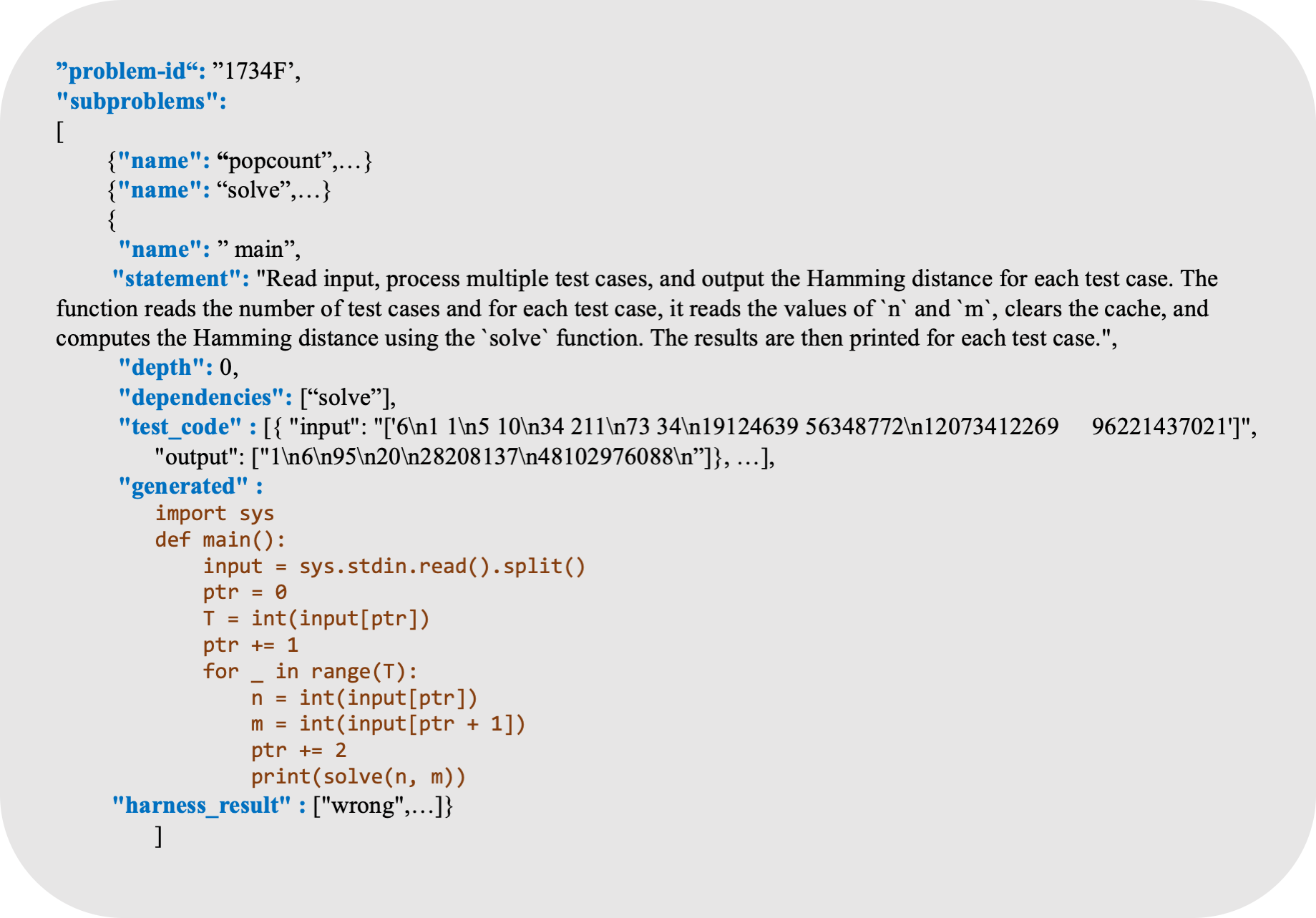}
    \caption{Example 2 of an Insufficient Globalization(IG) error by Deepseek-V3.The original problem is \href{https://codeforces.com/problemset/problem/1734/F}{1734F} Because the program reads new inputs and performs fresh calculations for each test case, failing to clear the cache beforehand can inadvertently reuse stale results from a previous case. This leads to incorrect output due to unintended interaction between caching and I/O.}
    \label{fig:IG2}
\end{figure*}

\begin{figure*}
    \centering
    \includegraphics[width=0.7\linewidth]{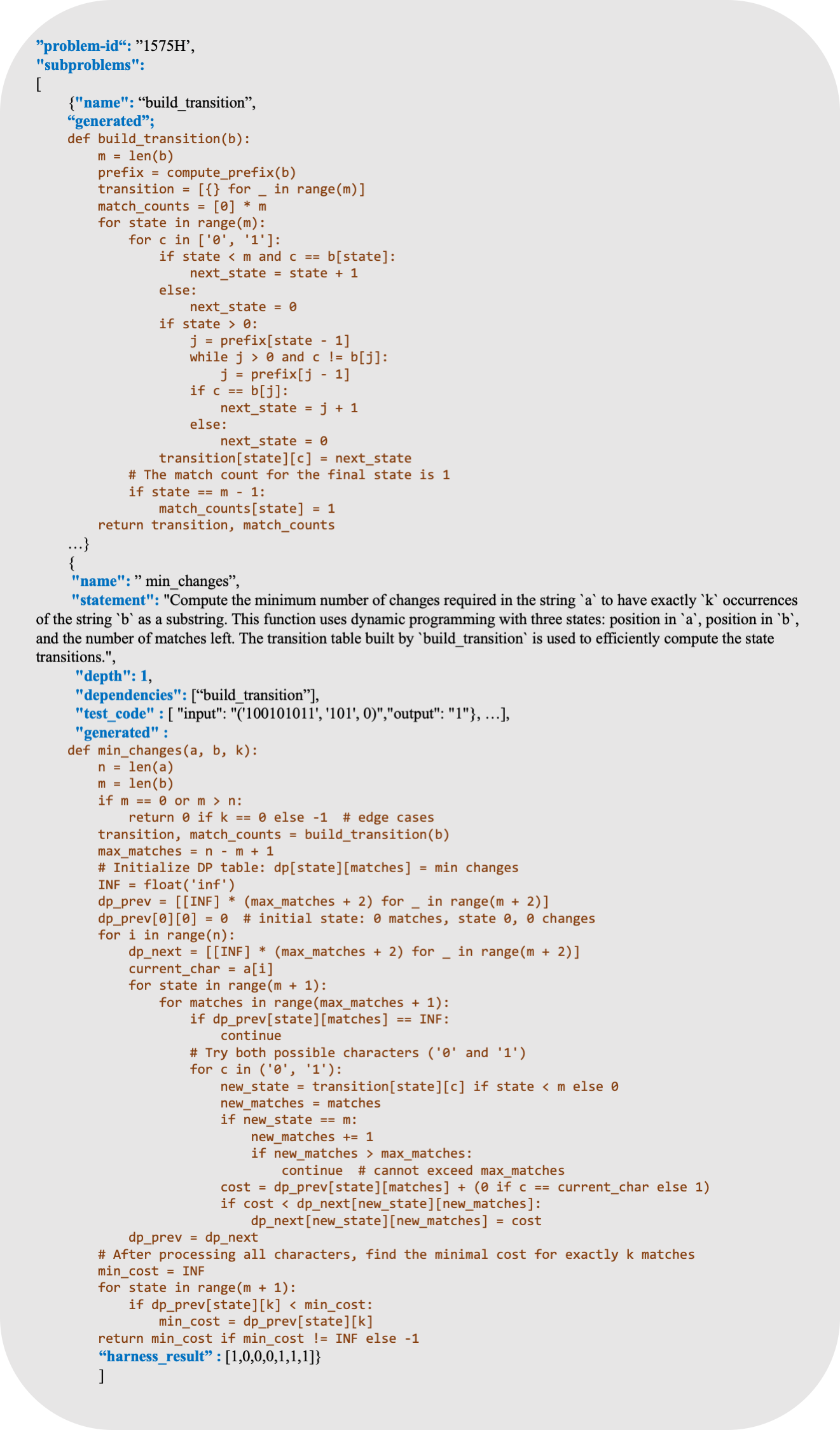}
    \caption{Example of an Instruction Misinterpretation(IM) error by Deepseek-V3.The original problem is \href{https://codeforces.com/problemset/problem/1575/H}{1575H} Although the DP routine correctly unpacks both transition and match\_counts from build\_transition(b), it merely treats match\_counts as an ordinary list. In reality, match\_counts carries two essential pieces of semantic information.(1)Match indicator.It denotes a complete match of b.(2)Backtrack hint. It's a hint in combination with the prefix function, indicating how far the automaton should jump back after a match to continue detecting overlapping occurrences. This error is caused by models' misinterpretation to the dependency relationship between tool function and top-level function}
    \label{fig:IM}
\end{figure*}
\section{Impact Statement}
\label{sec: Impact Statement}
This paper introduces CodeFlowBench, a comprehensive benchmark for evaluating code generation models in realistic multi-turn, dependency-driven development scenarios. For research, CodeFlowBench fills a critical gap by providing a standardized suite of tasks that require iterative reasoning, function dependency management, and end-to-end solution assembly. By exposing models’ deficiencies in global awareness, instruction consistency, and dependency handling, CodeFlowBench will catalyze the design of new architectures and training paradigms that explicitly model iterative workflows and cross-turn coherence. Its open dataset and evaluation protocol invite the community to develop and compare dependency-aware generation strategies, driving progress toward more robust and developer-friendly code assistants.

In industry, CodeFlowBench offers a practical yardstick for assessing the readiness of AI coding tools in real-world software development, where tasks rarely appear as isolated single-step prompts. Integrating CodeFlowBench into CI/CD pipelines can help organizations detect and remediate weaknesses in model-powered code suggestions before deployment, reducing debugging overhead and technical debt. By highlighting the importance of function reuse, import management, and state consistency across revisions, CodeFlowBench insights can inform best practices for AI-augmented coding workflows, accelerating adoption of reliable co-programming solutions. There are broader societal implications in enabling safer, more maintainable AI-generated code, yet none that we believe warrant special emphasis here.

\end{document}